\documentclass[a4paper,12pt,english]{article}

\pdfoutput=1

\usepackage[bindingoffset=0.3cm,textheight=22.5cm,hdivide={2cm,*,2cm}, vdivide={*,22cm,*}]{geometry}

\usepackage{amsmath,amsfonts,amssymb,babel,slashed,color,empheq,tensor}
\usepackage[svgnames]{xcolor}
\usepackage[pdftex,breaklinks,colorlinks=true,urlcolor=RoyalBlue,%
linkcolor=blue,citecolor=blue]{hyperref}

\usepackage[utf8,latin1]{inputenc}
\usepackage{graphicx}
\usepackage{dcolumn}

\usepackage{mathabx}
\usepackage{amsfonts}
\usepackage{latexsym}
\usepackage[normalem]{ulem}
\usepackage{mathrsfs}
\usepackage{cases}
\usepackage{bm}
\newcommand{\qm}[1]{``#1''}
\usepackage{hyperref}
\usepackage{stackengine,scalerel}
\hypersetup{colorlinks, linkcolor={red},citecolor={blue},urlcolor={blue}}  
\usepackage{bbold}
\usepackage{authblk}

\newcommand\overstar[1]{\ThisStyle{\ensurestackMath{%
  \setbox0=\hbox{$\SavedStyle#1$}%
  \stackengine{0pt}{\copy0}{\kern.2\ht0\smash{\SavedStyle*}}{O}{c}{F}{T}{S}}}}

\usepackage[numbers,square,comma,sort&compress]{natbib}
\usepackage{adjustbox}

\makeatletter

\def\R{{\mathbb R}} \def\C{{\mathbb C}} \def\N{{\mathbb N}}
 \def\one{\mbox{1 \kern-.59em {\rm l}}}

\newcommand{\tr}{\mathrm{tr}}
\newcommand{\End}{\mathrm{End}}
\def\hs{\mathfrak{hs}}

  \def\cC{{\cal C}} 
  \def\cF{{\cal F}} 
 \def\cH{{\cal H}}  
 \def\cK{{\cal K}} \def\cL{{\cal L}} 
\def\cM{{\cal M}}

 \def\a{\alpha}  \def\b{\beta}
 \def\g{\gamma} 
 \def\d{\delta} 

 \def\L{\Lambda}  
    \def\r{\rho}
\def\s{\sigma}  \def\t{\tau}

\newcommand{\eq}[1]{(\ref{#1})}
\newcommand{\del}{\partial}
\def\nn{\nonumber}

 \allowdisplaybreaks[3]

\makeatother

\renewcommand{\title}[1]{\vspace{10mm}\noindent{\Large{\bf

#1}}\vspace{8mm}} \newcommand{\authors}[1]{\noindent{\large

#1}\vspace{5mm}} \newcommand{\address}[1]{{\itshape #1\vspace{2mm}}}

\begin{document}

%%%% --- TITLE PAGE --- %%%%
% \begin{titlepage}
 \begin{flushright}
  UWThPh-2023-25 %\\
 \end{flushright}
\begin{center}
\title{ {\Large One-loop effective action of the IKKT model \\[1ex] for cosmological backgrounds} }

\vskip 3mm

\authors{Emmanuele Battista\footnote{emmanuele.battista@univie.ac.at, emmanuelebattista@gmail.com}, Harold C. Steinacker\footnote{harold.steinacker@univie.ac.at}}

 \vskip 2mm

  \address{ 
{\it Faculty of Physics, University of Vienna\\
 Boltzmanngasse 5, A-1090 Vienna, Austria  }  
   }

\bigskip

%\vskip 1.4cm

%%%% --- ABSTRACT --- %%%%
\textbf{Abstract}
\vskip 3mm

\begin{minipage}{14.3cm}%

We study cosmological solutions of the IKKT model with  $k=-1$ FLWR geometry, taking into account one-loop corrections. A previously discussed covariant quantum spacetime is found to be stabilized through one-loop effects at early times, without adding a mass term to the model. At late times, this background is modified
and approaches a solution of the classical model where $a(t) \sim const$, but the dilaton  decreases in time.
This suggests that a more complete treatment of the system is required in the late-time regime.

\end{minipage}

\end{center}

{
  \hypersetup{linkcolor=black}
  \tableofcontents
}

\section{Introduction}

One of the major problems of theoretical physics consists in providing a unified framework for quantum mechanics and general relativity. Among the various proposed approaches, such as loop quantum gravity \cite{Rovelli1997,Smolin2004,Ashtekar2021}, effective field theories of gravity \citep{Donoghue1994,Donoghue2017a,Burgess2020,Donoghue2022} (see e.g. Refs. \cite{Afrin2022,Brahma2020,Zhu2020,Bojowald2011,Battista2014a,Battista2014b,Bjerrum-Bohr2014,Battista2015a,Battista2015b,Bjerrum-Bohr2016,Battista2017a,Mandal2022} for their applications and potential observational tests) and string theory, the Ishibashi-Kawai-Kitazawa-Tsuchiya (IKKT) matrix model has been proposed as a constructive definition of one of the corners of string theory \cite{Ishibashi1996}. 
This framework leads to a new approach to obtain cosmological spacetimes.
In particular, solutions  have been found \cite{Steinacker:2017vqw,Steinacker:2017bhb,Sperling:2019xar}, which can be interpreted as 3+1 dimensional quantum geometries describing an effective Friedmann-Lema\^{i}tre-Robertson-Walker (FLRW) cosmology with a big bounce. We have analyzed the main features of the underlying bouncing cosmological model including the propagation of  bosonic and fermionic fields in Refs. \cite{Battista2022a,Battista:2023kcd,Battista:2022vvl}. For other approaches to bouncing cosmology we refer the reader to Refs. \cite{Escofet2015,Cai2016,Cai2017,Klinkhamer2019b,Ilyas2020,Odintsov2020a,Battista2020a,Zhu2021,Wang2021a,Zhu2023} and references therein, 
and to Refs. \cite{Brahma:2021tkh,Brahma:2022ikl}
for recent works on cosmology from matrix theory from a somewhat different angle.

The IKKT matrix model \cite{Ishibashi1996} is defined by the action
\begin{align}
S[T,\Psi] = \frac{1}{g^2}{\rm Tr}\big( [T^A,T^B][T_{A},T_{B}] 
\,\, + \overline\Psi \Gamma_A[T^A,\Psi] \big) \ ,
\label{MM-action-IKKT}
\end{align}
where  $T^A$ $ (A = 0,..., 9)$ are hermitian matrices and $\Psi$  
Grassmann-matrix-valued Majorana-Weyl spinors of $SO(9,1)$. The model is uniquely determined by maximal supersymmetry, which is essential 
to obtain a sufficiently local quantum effective action on 3+1 dimensional background branes.

Due to its special properties, the IKKT model shares the special status of superstring string theory as an approach to incorporate gravity into a consistent quantum theory. Remarkably, it offers a novel mechanism to obtain gravity in 3+1 dimensions, which is distinct from the standard mechanism 
considered in string theory leading to 9+1 dimensional gravity. This new mechanism arises in the weak coupling regime on 3+1 dimensional branes at one loop, where the fluctuation modes propagate on the brane, and do not escape into the bulk. 
More specifically, it was recently shown  \cite{Steinacker:2021yxt,Steinacker:2023myp} that the Einstein-Hilbert action arises in the one-loop effective action on suitable backgrounds. The mechanism 
requires 3+1 dimensional spacetime branes with a certain 
product structure,
without compactifying target space.
Therefore the issue of the string theory landscape does not arise. However, it remains to be shown that this \qm{emergent} gravity arising on the 3+1 dimensional spacetime brane can be (near-) realistic.

In the present paper, we address this question for the particular case of the cosmological FLRW solution given in Ref. \cite{Sperling:2019xar}, and study its properties at one loop.
We will study the one-loop dynamics of backgrounds
$T^A$ which can be interpreted in the semiclassical regime 
as FLRW cosmological spacetime.
These  backgrounds can then be described as symplectic manifolds $\cM$ embedded in target space via the matrices
\begin{align}
   T^A\ \sim t^A: \ \ \cM \hookrightarrow \R^{9,1} \ .
   \label{background}
\end{align}
We will restrict ourselves to the semiclassical regime of the geometry in this paper, where all matrices are replaced by functions, and commutators by Poisson brackets, i.e.,  $[.,.] \sim i \{.,.\}$.
Moreover, we consider only 3+1 dimensional spacetime branes $\cM^{3,1}$ embedded along the first 4 matrix directions labeled by 
 $a=0,..,3$.  However these will be accompanied by fuzzy compact branes $\cK$ in transversal directions $i=4,...,9$, which is essential to obtain the 
Einstein-Hilbert action at one loop.
 An introduction and motivation for this framework can be found in Refs. \cite{Steinacker:2010rh,Steinacker:2019fcb}, see also e.g. Refs. \cite{Chaney:2015ktw,Steinacker:2017bhb,Sperling:2019xar,Steinacker:2021yxt,Brahma:2022dsd,Karczmarek:2022ejn,Battista:2022vvl} for related work in this context, as well as Refs. \cite{Nishimura:2019qal,Nishimura:2022alt,Anagnostopoulos:2022dak} for efforts towards establishing the emergence of 3+1 large dimensions in numerical simulations.

\emph{Notations.} We use metric signature  $(-,+,+,+)$ and units $G=c=\hbar=1$. $\alpha,\beta,\dots=0,\dots,3$ are coordinate indices, whereas $a,b,\dots=0,\dots,3$  are tetrad indices. The flat metric is indicated by $\eta^{ab }=\eta_{ab }={\rm diag}(-1,1,1,1)$.

\section{Geometrical structures}

In this section, we recall some geometric structures relevant to the present framework, and 
study properties of classical divergence-free frames, ignoring possible higher spin ($\hs$) contributions for now. These arise generically on the class of backgrounds due to a hidden internal $S^2$ \cite{Sperling:2019xar}, and 
will be taken into account in section \ref{sec:frame-reconstruct}.

The fundamental geometrical object in the matrix model is the frame \cite{Steinacker:2020xph}
\begin{align}
\tensor{E}{^{ a}^\mu} &:= \{T^{ a},x^\mu\}  \ ,
 \label{frame-general}
\end{align}
which arises as Hamiltonian vector fields generated by the matrix background $T^a$.
The metric is related to the frame through
\begin{align}
 G^{\mu\nu} = \r^{-2}\, \g^{\mu\nu} , \qquad
 \g^{\mu\nu} = \tensor{E}{_{ a}^\mu} \tensor{E}{^{ a}^\nu} \  ,
  \label{eff-metric-def}
\end{align}
where the dilaton $\r$ relates the symplectic density
$\r_M$  on $\cM^{3,1}$  to the Riemannian density
$\sqrt{|G|}$  via (with $G:= \det G_{\mu \nu}$ and $\gamma:= \det \gamma_{\mu \nu}$)  
\begin{align}
\r_M  = \r^{-2}\sqrt{|G|} \ = \ \r^2 \sqrt{|\g|} \ .
\label{rho-M-G-relation}
\end{align}
In Cartesian coordinates, the symplectic volume form is given explicitly by
$\Omega = \rho_M d^4 x$ with\footnote{Strictly speaking $\Omega$ arises from the symplectic form on the underlying 6-dimensional space $\C P^{1,2}$, which is an $S^2$ fiber over $\cM^{3,1}$. The $S^2$ factor is suppressed here.}
\begin{align}
    \r_M = \frac{1}{\sinh \eta} \ .
    \label{rho-M-cart}
\end{align}
Since symplectic manifolds are rigid, this can also be used for deformed backgrounds.

Considering that the frame arises in the present framework as symplectic vector field, it must satisfy the following divergence constraint \cite{Fredenhagen2021}:
\begin{align}
 \nabla^{(G)}_\mu(\r^{-2}\tensor{E}{^{  a}^\mu}) &= 0
 =  \del_\mu(\r_M\tensor{E}{^{  a}^\mu}) 
  \label{div-constraint-E}
\end{align}
in local coordinates on $\cM$.  

\paragraph{Weitzenb\"ock connection and torsion.}

In the framework of matrix models, 
the Weitzenb\"ock connection associated to that frame  \eqref{frame-general}  turns out to be useful. This connection is defined by $\nabla E^a = 0$, with trivial curvature but nontrivial torsion. This is natural because the torsion with two frame indices satisfies \cite{Sperling:2019xar}
\begin{align}
    T^{ab \mu} &= \{\cF^{ab},x^\mu\} 
    = - E^{a\nu}\del_\nu E^{b\mu} + E^{b\nu}\del_\nu E^{a\mu}\ ,
    \label{torsion-F-identity}
\end{align}
where 
\begin{align}
    \cF^{ab} = -\{T^a,T^b\}
\end{align}
is the \qm{field strength} of the background $T^a$.
Recasting $T^{ab \mu}$ as a covariant tensor using the frame, the following contraction identities hold:
 \begin{align}
   \tensor{T}{_\mu_{\s}^\mu} = \tensor{K}{_\mu_{\s}^\mu} &= 2 \r^{-1}\del_\s\r \ ,
 \label{tilde-T-T-contract}
 \end{align}
where $\tensor{T}{_\mu_{\s}^\nu} = \tensor{\Gamma}{_\mu_{\s}^\nu}-\tensor{\Gamma}{_{\s}_\mu^\nu}$ and $\tensor{ K}{_{\mu}_{\nu}^{\s}}=\frac{1}{2} \left(\tensor{ T}{_{\mu}_{\nu}^{\s}}+\tensor{T}{^\sigma_\mu_\nu} -\tensor{T}{_\nu^\sigma_\mu}\right)$ are the torsion and contorsion tensors, respectively.

\subsection{The undeformed FLRW background $\cM^{3,1}$}

We start with the undeformed background considered in Ref. \cite{Sperling:2019xar}, which is defined by the matrix configuration 
\begin{align}
    \bar T^a := \frac 1R\, \cM^{a4} \sim t^a
\label{undeformed-background}
\end{align}
considered as a function on the underlying 6-dimensional bundle space $\C P^{1,2}$, which is a coadjoint orbit of $SO(4,2)$ with generators $\cM^{ab}$ (here $a,b = 0,...,5$). 
Locally, this bundle has the structure $\C P^{1,2} \cong \cM^{3,1} \times S^2$, where 
the spacetime manifold $\cM^{3,1}$ is described by Cartesian coordinates $x^\mu$, and the internal sphere by the $t^\nu$. These 
satisfy the following relations  \cite{Sperling:2019xar}:
\begin{subequations}
\label{relations-Cartesian}
\begin{align}
    x^\mu x^\nu \eta_{\mu\nu} &= - R^2 \cosh^2\eta,
    \label{x-mu-x-mu-prod}\\
    t^\mu t^\nu \eta_{\mu\nu} &= \tilde R^{-2} \cosh^2\eta,
    \label{t-mu-t-mu}\\
     x^\mu t^\nu \eta_{\mu\nu} &= 0.
     \label{x-mu-t-mu}
\end{align}    
\end{subequations}
They generate the algebra of functions on $\C P^{1,2}$, which can accordingly be viewed as algebra of higher-spin $\hs$ valued functions on $\cM^{3,1}$, which decomposes into the spin $s$ sectors $\cC^s$ spanned by (irreducible) polynomials of order $s$ in $t^\mu$.

We also recall the following identity which relates $\cM^{3,1}$ with a hyperboloid $H^4 \subset \R^{4,1}$
\begin{align}
x_4^2 = - x_\mu x^\mu - R^2,
\label{x4-squared}
\end{align}
thereby defining $x^4$. This entails 
\begin{align}
%   x^4 \frac{\del}{\del x^4} = - x^\mu \del_\mu
 x^4\frac{\del x^4}{\del x^\mu} = - x^\mu \ .
\end{align}
For more details we refer the reader to Ref. \cite{Sperling:2019xar}.

\subsection{$SO(3)$-invariant time-dependent frames }

In order to describe more general $SO(3)$-invariant time-dependent frames, we need to generalize some results of Ref. \cite{Fredenhagen2021}. Using Cartesian coordinates $x^\mu$ and introducing the notation
\begin{align}
\tilde t   &= x^0,
\nn  \\ 
r^2 &= \delta_{ij} x^i x^j, 
\end{align}
(here the time parameter $\tilde t$ should not be confused with the $t^\mu$ generators in Eq. \eqref{undeformed-background} ff.; moreover, Latin indices $i,j,\dots=1,2,3$),
the most general spherically symmetric frame $\tensor{E}{^a_\mu}$ can be written as
\begin{align}
\tensor{E}{^0_0} &= A\ ,\nonumber\\
\tensor{E}{^i_0} &= Ex^{i}\ ,\nonumber\\
\tensor{E}{^0_i} &= Dx^{i}\ ,\nonumber\\
\tensor{E}{^i_j} &= Fx^{i}x^{j} + \delta^{i}_{\ j}B + S\epsilon_{ijm} x^{m} \ 
\label{ansatz}
\end{align}
in Cartesian coordinates.
It is simple to show that $D$ and $F$ can be made to vanish via a   change of coordinates \cite{Fredenhagen2021}; moreover, the totally antisymmetric part of the torsion tensor, which is related to the contribution of the axion, is zero if $S=0$. With these assumptions, the inverse frame $\tensor{E}{_a^\mu}$ is
\begin{align}
\tensor{E}{_0^0} &= A^{-1}\ ,\nonumber\\
\tensor{E}{_i^0} &= 0\ ,\nonumber\\
\tensor{E}{_0^i} &= -\frac{E}{AB} x^i\ ,\nonumber\\
\tensor{E}{_i^j} &= \delta^{i}_{\ j}B^{-1}  \ .
\label{inverse-frame}
\end{align}

Bearing in mind Eqs. \eqref{ansatz} and \eqref{inverse-frame},  the relation \eqref{tilde-T-T-contract} governing the behaviour of the  dilaton $\rho$ yields 
\begin{align}
 -\frac{2}{\rho}\partial_{\mu}\rho =
\tensor{T}{_\mu_\nu_a}\tensor{E}{^a^\nu} 
= \partial_{\mu}\tensor{E}{^0_0}\tensor{E}{_0^0} 
+ \partial_{\mu}\tensor{E}{^i_j}\tensor{E}{_i^j} 
- \partial_{i}\tensor{E}{^0_\mu}\tensor{E}{_0^i} - \partial_{j}\tensor{E}{^i_\mu}\tensor{E}{_i^j}-\partial_{0}\tensor{E}{^0_\mu}\tensor{E}{_0^0}  \ .
\label{T-contract-2}
\end{align}
The spacelike component $\mu=k$ of the above equations leads to
\begin{align}
 -\frac{2}{\rho}\partial_{k}\rho = \frac{1}{A} \partial_{k}A + \frac{2}{B} \partial_{k}B \ ,    
\end{align}
whose solution is
\begin{align}
\rho^2 = \frac{f\left(\tilde t \right)}{AB^2} \ , 
\label{rho-squared-sol}
\end{align}
$f\left(\tilde t \right)$ being an arbitrary function of $\tilde t$. On the other hand, the timelike component $\mu=0$ of Eq. \eqref{T-contract-2} gives
\begin{align}
\frac{\dot f}{f} - \frac{\dot A}{A} + \frac{\dot B}{B} + \frac{1}{B} \left[ \frac{E r \partial_r A }{A} - \left(r \partial_r E + 3E\right) \right]=0 \ ,   
\label{timelike-eq-frame}
\end{align}
where we have exploited Eq. \eqref{rho-squared-sol}, and  the dot stands for differentiation with respect to  $\tilde t$. The solution of Eq. \eqref{timelike-eq-frame} can be easily obtained if we assume the simplifying hypothesis $E=0$. Indeed, in this case we get
\begin{align}
 B= \frac{A}{f\left(\tilde t \right)} g\left(r \right),
 \label{B-coeff-sol}
\end{align}
where $g\left(r \right)$ is an arbitrary function of $r$. It follows from Eqs. \eqref{rho-squared-sol} and \eqref{B-coeff-sol} that the divergence constraint (see Eq. \eqref{div-constraint-E})  is automatically fulfilled independently of the form assumed by $f\left(\tilde t \right)$ and  $g\left(r \right)$. This means that we can consistently put $f\left(\tilde t \right)=1=g\left(r \right)$, which owing to Eq. \eqref{B-coeff-sol} implies $A=B$. Then the effective metric $G_{\mu \nu}$ becomes finally (cf. Eq. \eqref{eff-metric-def})  
\begin{align}
ds^2_G = -\frac{1}{A} d \tilde{t}^2 + \frac{1}{A} \delta_{ij} dx^i d x^j,  
\end{align}
which provides the $SO(3,1)$-invariant geometry considered in the following. More generally, we can set $g\left(r \right) = 1$ but leave $f\left(\tilde t \right)$ arbitrary. Then we get 
\begin{align}
ds^2_G = -\frac{1}{A} d t^{'2} + \frac{f}{A} \delta_{ij} dx^i d x^j,     
\end{align}
where
\begin{align}
  d t^{'2} =f\left(\tilde t \right)^3  d \tilde{t}^2. 
\end{align}

\subsection{$SO(3,1)$-invariant frames}
\label{sec:so31-frames}

To obtain a FLRW cosmology with $k=-1$ geometry,
 it is natural to require that not only the metric but also the frame is invariant under the $SO(3,1)$ isometry group. This isometry should naturally act on the frame indices as 
\begin{align}
\tensor{E}{^{ a}^\mu} \to 
\L^{ a}_{ b}
\L^{\mu}_{\nu} \tensor{E}{^{ b}^\nu},
\end{align}
in Cartesian coordinates.
Then  $SO(3,1)$  invariance implies that 
\begin{align}
\tensor{E}{^{ a}_\mu} = A(\eta)\big(\delta^a_\mu + F(\eta) x^a x_\mu\big),
\label{frame-ansatz-F}
\end{align}
or equivalently 
\begin{align}
\tensor{E}{^{ a}^\mu} = \frac 1{A(\eta)}\big(\eta^{a\mu} - \tilde F(\eta) x^a x^\mu\big) \ .
\end{align}
Here we define $\eta$ through (cf. \eqref{relations-Cartesian})
\begin{align}
x^2_4 := R^2 \sinh^2(\eta) := -R^2 + x_0^2 - \sum x_i^2,
\label{constraint-H4}
\end{align}
which is a measure for the FLRW time, labeling the spacelike hypberboloids $H^3$.
Again, we can remove the $F$ term using a suitable change of variables $\tilde x^\mu = f(\eta) x^\mu$, so that (hereafter, a prime indicates the derivative with respect to $x^4$) 
\begin{align}
 E^a =  \tensor{E}{^{ a}_\mu} d x^\mu 
  &= \tensor{\tilde E}{^{ a}_\mu} d\tilde x^\mu 
   = \tensor{\tilde E}{^{ a}_\mu} (f d x^\mu + f' x^\mu dx^4) \nn\\
    &= \tensor{\tilde E}{^{ a}_\mu} (f\d^\mu_\nu - \frac 1{x^4} f' x^\mu x_\nu) dx^\nu \nn\\
    &= \tilde A(\eta)(f \delta^a_\nu - \frac 1{x^4 } f' x^a x_\nu) dx^\nu 
\end{align}
with
\begin{align}
\tensor{\tilde E}{^{ a}_\mu} = \tilde A(\eta) \delta^a_\mu \ .  
\label{Frame-and-A}
\end{align}
    This holds if the following two conditions are satisfied 
\begin{align}
    f \tilde A(\eta) &= A(\eta) \ , \nn\\
    \tilde A(\eta) \frac 1{x^4} f' &= - A F \ ,
\end{align}
which determines $f$ through
\begin{align}
  \frac 1{x^4 f} f' &= - F \ .
  \label{elimination-xx-term}
\end{align}
We will accordingly assume that $F=0$ in the adapted coordinates. Then
the divergence constraint \eq{div-constraint-E} implies 
\begin{align}
    A(\eta) = c \, \r_M 
    \label{A-rhoM-div}
\end{align}
where $c$ is an arbitrary constant that will be set equal to one. However, $\rho_M = \rho_M(\eta)$ should then be considered as undetermined.
The dilaton $\rho$ is determined by the contraction identity \eq{tilde-T-T-contract},
which similarly as before (cf. Eq. \eqref{rho-squared-sol}) yields
\begin{align}
\label{rho-equation}
    \r^2 = A^{-3} =  \r_M^{-3} \ .
\end{align}
Then the effective metric is 
\begin{align}
G_{\mu\nu} = \frac 1{A(\eta)} \eta_{\mu\nu} = \frac 1{\r_M(\eta)} \eta_{\mu\nu} \  .
\label{eff-metric-and-A}
\end{align}
This provides the most general FLRW metric with $k=-1$. 
That metric is simply obtained by setting $F=0$
in the above ansatz \eqref{frame-ansatz-F}, which will be assumed henceforth.

%***
%which means that we end up precisely with the standard metric \eq{} on $\cM^{3,1 }$. This is very strange: it means that $\cM^{3,1}$ is a preferred FLRW cosmology on general structural grounds in the present framework, independent of the dynamics or matter content.
%How can this be?? (Note that $m_\cK$ is still dynamical, that might make up for it ... but no...)

%Of course we can consider frames which are not invariant under $SO(3,1)$; but then presumably the torsion would not be $SO(3,1)$ invariant.

It follows from Eq. \eqref{eff-metric-and-A} and the hyperbolic  parametrization \cite{Battista2022a}
\begin{align}
 \begin{pmatrix}
  x^0 \\ x^1 \\ x^2 \\x^3 
 \end{pmatrix}
= R \cosh\eta 
\begin{pmatrix}
\cosh\chi \\
\sinh\chi \, \sin\theta \, \cos\varphi \\
\sinh\chi \, \sin\theta \, \sin\varphi \\
\sinh\chi \, \cos\theta
\end{pmatrix} \ ,
\label{embedding-3d-hyperboloid}
\end{align} 
that the effective metric can be written as 
\begin{align}
 d s^2_G = G_{\mu\nu} d x^\mu d x^\nu 
&= -R^2 A^{-1} \sinh^2\eta \,   d \eta^2 + R^2 A^{-1} \cosh^2\eta\, d \Sigma^2,
\label{eff-metric-FRW}
\end{align}
where 
\begin{align}
    d\Sigma^2 = d\chi^2 + \sinh^2\chi (d\theta^2 + \sin^2 \theta \, d\varphi^2),
    \label{dSigma2}
\end{align}
is the invariant length element on the spacelike hyperboloids $H^3$. We can bring Eq. \eqref{eff-metric-FRW} to the standard FLRW form 
\begin{align}
 d s^2_G = -d t^2 + a(t)^2 d\Sigma^2
 \label{FRW-standard-metric}
\end{align}
via the relations
\begin{align}
dt^2 &= R^2 A^{-1} \sinh^2\eta \,   d \eta^2,
\label{dt-d-eta}
\\
a(t)^2 &=   R^2 A^{-1} \cosh^2\eta \ .  
\label{relation-a-t-eta}
\end{align}
This leads to a coasting late-time evolution  \cite{Sperling:2019xar} with
\begin{align}
    a(t) \sim \frac 32 t \ .
\end{align}
The Weitzenb\"ock connection associated to the frame  \eqref{Frame-and-A} reads as
\begin{align}
\tensor{\Gamma}{_\nu_\rho^\mu} = - \tensor{E}{^a_\rho} \partial_\nu \tensor{E}{_a^\mu}= -\frac{\partial_\eta A}{AR^2 \sinh \eta \cosh \eta}  x_\nu \delta^\mu_\rho ,
\end{align}
where   we have exploited the relation \cite{Sperling:2019xar}
\begin{align}
 \partial_\mu f(\eta) = -\frac{x_\mu}{R^2 \sinh \eta \cosh \eta}  \partial_\eta  f(\eta) .
 \label{time-derivative-cartes}
\end{align}
Therefore, the torsion tensor is formally
\begin{align}
 \tensor{T}{_\nu_{\rho}^\mu} = \tensor{\Gamma}{_\nu_\rho^\mu} -\tensor{\Gamma}{_\rho_\nu^\mu}  = \frac{\partial_\eta A}{AR^2 \sinh \eta \cosh \eta}  \left(x_\rho \delta^\mu_\nu - x_\nu \delta^\mu_\rho\right) .
 \label{torsion-expr}
\end{align}
However,  it should be kept in mind that in the presence of nontrivial $A(\eta)$, the torsion typically acquires $\hs$ valued components which may not be negligible.

\section{Generalized $k=-1$ FLRW matrix background}
\label{sec:frame-reconstruct}

So far, we considered the cosmological frames which may arise in the matrix model setting. However,  these frames need to be implemented 
through generators $T^a$ via Poisson brackets $\tensor{E}{^{ a}^\mu} := \{T^{ a},x^\mu\}$ (cf. Eq. \eqref{frame-general}). For the undeformed solution $\cM^{3,1}$ \cite{Sperling:2019xar} of the IKKT model with mass term, this background is given by 
\begin{align}
T^a &= t^a \ 
\end{align}
leading to the frame
\begin{align}
    E^{a\mu} = \{T^a,x^\mu\} = A^{-1} \eta^{a\mu}, \qquad A = \frac 1{\sinh(\eta)} \ .
\end{align}
Requiring $SO(3,1)$ invariance strongly suggests to consider the following generalized $SO(3,1)$-covariant  
matrix background 
\begin{align}
T^a &= \a(\eta) t^a + \b(\eta) x^a,  
\label{FLRW-frame--1}
\end{align}
which will lead to homogeneous and isotropic
cosmological backgrounds
corresponding to a FLRW geometry with
$k=-1$.
As shown  in appendix \ref{sec:app-normal-gauge}, it is always possible to
use  a $SO(3,1)$-invariant gauge transformation $\d_\L = \{\L(x^4),.\}$ to reduce this background to the form
\begin{align}
\boxed{
    \ T^a = \a(\eta) t^a, \ 
    }
    \label{FLRW-background-alpha}
\end{align}
denoted as standard FLRW gauge.
Note that all $SO(3,1)$-invariant functions on the underlying background are functions of
\begin{align}
x^4= R\sinh \eta \ .
\label{x4-coord}
\end{align}
To compute the resulting frame, we note that Poisson brackets of 
$\eta=\eta(x^4)$ can be evaluated as follows
\begin{align}
 \{x^4,t^\mu\} &= R^{-1} x^\mu,   \nn\\
 \{x^4,x^\mu\} &= \tilde R^2 R t^\mu  \ .
\end{align}
Then the frame is found to be
\begin{align}
    E^{a\mu} = \{T^a,x^\mu\} 
    &= \sinh\eta \, \a \eta^{a\mu} 
+ \tilde R^2 R  \a' t^a t^\mu.
\label{frame-FLRW-hs-x}
\end{align}
As shown in appendix \ref{sec:loc-normal-coords},
these $\hs$ components proportional to $ t^a t^\mu$ can be eliminated locally in generalized normal coordinates, at the expense of manifest  $SO(3,1)$ invariance. The resulting framework of \qm{higher-spin geometry} is still somewhat obscure, and we shall compute what appears to be the most reasonable effective metric in section \ref{sec:FLRW-one-loop} and Appendix \ref{sec:loc-normal-coords}.
However as a first and more conservative step, we  accept the presence of small $\cC^2$ components of the frame in this section, and  demand that these are negligible in some \qm{linearized} or \qm{weak-gravity} regime.

\subsection{Linearized regime}
\label{sec:linearized}

The $\cC^0$ and  $\cC^2$ components of the frame
can be separated by projecting or averaging  over the $S^2$ fiber, which is achieved using \cite{Sperling:2019xar} 
\begin{align}
[t^a t^\mu]_0 &= \frac{1}{3\tilde R^2} \big( \eta^{a\mu} \cosh^2\eta + R^{-2}\, x^a x^\mu   \big).
\label{t-mu-t-a-spin-zero}    
\end{align}
Then the classical and $\cC^2$-valued components of the  effective  frame are 
\begin{align}
    e^{a\mu} = [E^{a\mu}]_0 
    &= \big( \a \sinh\eta
    + \frac{1}{3} R  \a' \cosh^2\eta \big)\eta^{a\mu}
    + \frac{1}{3}  \a' \, R^{-1} \, x^a x^\mu,  
    \nn\\
    [E^{a\mu}]_2 &=  \tilde R^2 R  \a' [t^a t^\mu]_2 \ \  = O(R  \a' \cosh^2\eta)\ .
\end{align}
The above projection $[.]_0$ to $\cC^0$ is  justified in the weak-gravity regime, which means that the $\cC^2$ components of the frame are much smaller than the $\cC^0$ components:
\begin{align}
      [E^{a\mu}]_0   & \gg [E^{a\mu}]_2 \ .
\end{align}
This boils down to 
\begin{align}
    \a  \sinh\eta &\gg   R \a' \cosh^2\eta \ ,
    %\nn\\
   %\sinh(\eta)  &\gg  R \frac{\a'}{\a} \cosh^2(\eta) 
\end{align}
i.e.
\begin{align}
 \varepsilon := 
 \frac{1}{\a}\frac{d\a}{d\eta}
= R \frac{\a'}{\a} \cosh\eta   \ll 1 \ .
  \label{lin-approx-cond-nohs}
\end{align}
Here the parameter $\varepsilon$ measures the quality of the classical approximation (neglecting the $\hs$ components),
denoted as \qm{slow-rolling} approximation henceforth.
However, $\varepsilon \ll 1$ would imply 
\begin{align} \label{tetrad-with-alpha-1}
e^{a\mu} = [E^{a\mu}]_0 \approx E^{a\mu}  \approx \a \, \eta^{a\mu} \sinh\eta \ ,
\end{align}
which would imply that the cosmology is  close to the unperturbed one with $\a = const$.
To see this, consider the cosmic scale factor $a(t)$  obtained in the late-time regime $\eta\gg 1$ using Eq. 
\eqref{relation-a-t-eta}
 as
\begin{align}
 a(t)^2 \approx R^2 \alpha \cosh^2 \eta \sinh \eta \ ,  
 \label{a-squared-first-relation}
\end{align}
which gives (hereafter a dot denotes the  time derivative with respect to $t$, e.g., $\dot a = \frac{d}{dt} a(t)$)
\begin{align}
\frac{\dot a}{a} & = \frac{1}{2} \left[\frac{\dot \alpha}{\alpha} + \frac{1}{R}\sqrt{\frac{A}{\sinh^2 \eta}} \left(2 \tanh \eta + \coth \eta\right)\right],  
\label{a-dot-over-a}
%\frac{a'}{a}&=  \frac{1}{2\alpha}\frac{1}{R \cosh  \eta } \left[\partial_\eta \alpha +\alpha \left(2 \tanh \eta + \coth \eta\right)\right],
\end{align}
where 
\begin{align}
A= \frac 1{\alpha \sinh \eta}  
\label{A-expression}
\end{align}
(cf. Eq. \eqref{tetrad-with-alpha-1}) and we have  used Eq. \eqref{dt-d-eta}. This can be rewritten using 
\begin{align}
\frac{\dot \alpha}{\alpha} &=  \frac{\alpha'}{\alpha} \dot{x}^4 = \frac{\alpha'}{\alpha}  (R \cosh\eta) \dot{\eta} 
= \frac{\alpha'}{\alpha} \cosh \eta \sqrt{\frac{A}{\sinh^2\eta}},
\end{align}
where Eq. \eqref{dt-d-eta} has been employed again. 
%\begin{align}
%\frac{\dot a}{a} &=  \frac{a'}{a} \frac{d x^4}{dt} =\frac{\alpha'}{\alpha}() = ...
%\end{align}
At late times $\eta\gg 1$, this simplifies as 
\begin{align}
    \frac{\dot \alpha}{\alpha} &\approx \frac{\alpha'}{\alpha}\sqrt{A},
\end{align}
while Eq. \eqref{a-dot-over-a} gives the Hubble rate
\begin{align}
H := \frac{\dot a}{a} &\approx  \frac{1}{2} \left[\frac{\dot \alpha}{\alpha} + \frac{3}{R}\sqrt{\frac{A}{\sinh^2 \eta}} \right] \nn\\
 &\approx 
 %\frac{1}{2} \left[\frac{\alpha'}{\alpha} + \frac{3}{R\sinh\eta} %\right] \sqrt{A} \nn\\
 % &= 
 \frac{1}{2} \left[\frac{\alpha'}{\alpha}R\sinh\eta + 3 \right] \frac{\sqrt{A}}{R\sinh\eta}. 
\end{align}
This means that the condition $\varepsilon \ll 1$ (cf. Eq. \eqref{lin-approx-cond-nohs}) is valid if and only if
\begin{align}
    \frac{H}{\bar H} \approx 1
\end{align}
where 
\begin{align}
\bar H &= \frac 32 \frac{\sqrt{A}}{R\sinh\eta} 
\approx \frac 32 \frac 1{\bar a(t)},   
\\
\bar a(t) &= R \sqrt{\alpha \cosh^2 \eta \sinh \eta} 
\approx \frac {R \cosh\eta}{\sqrt{A}}
\label{bar-a}
\end{align}
 are the Hubble rate and cosmic scale parameter
 for the \qm{slow-rolling} approximation $\a=const$.
 More explicitly, then the slow-rolling condition becomes 
 \begin{align}
     \varepsilon = R \frac{\a'}{\a} \cosh\eta
     \approx \frac{\dot \a}{\a}  \frac{R \cosh\eta }{\sqrt{A}} 
      =  \frac{\dot \a}{\a} \bar a(t) &\ll 1,   
 \end{align}
      hence
 \begin{align}      
       \frac{\dot \a}{\a} &\ll \frac 1{\bar a(t)} \sim O\big(t^{-1}), 
 \end{align}
 which means that the cosmic evolution is  close to the undeformed background geometry 
$\cM^{3,1}$ with $a(t) \sim \frac 32 t$. Therefore, to describe significantly different cosmic time evolutions, one needs to keep the $\hs$ components.  This will be done in the following.

%Maybe it is better to estimate $\varepsilon$ by the cosmic acceleration
%\begin{align}
%    \frac{\ddot a}{a} &= 
%\end{align}
%or the deceleration parameter 
%\begin{equation}
%    q = \frac{\ddot a a}{\dot a^2} = ...
%\end{equation}

%Relate this to the curvature and torsion below. 

%Can we relate this to the weak gravity regime?

\subsection{$\hs$-valued  inverse tetrad and  metric }

In order to evaluate the metric $G_{\mu \nu}$, we first need to work out the tetrad frame components $\tensor{E}{^a_\mu}$ in the presence of $\hs$ components. Starting from Eq. \eqref{frame-FLRW-hs-x} and exploiting the relation $\tensor{E}{^a_\mu} E^{b \mu} = \eta^{ab}$ , we find
\begin{align}
\tensor{E}{^a_\mu} = \frac{1}{\alpha \sinh \eta} \left(\delta^a_\mu - t^a t_\mu \mathscr{D}\right),    
\end{align}
which yields (cf. Eq. \eqref{eff-metric-def})
\begin{align} 
 G_{\mu \nu} = \rho^2  \tensor{E}{^a_\mu} \tensor{E}{^b_\nu} \eta_{ab} = \frac{\rho^2}{\alpha^2 \sinh^2 \eta} \left(\eta_{\mu \nu} + \tilde{\mathscr{D}} t_\mu t_\nu\right) = \rho^2 \gamma_{\mu \nu} \ ,
 \label{eff-metric-computation}
\end{align}
where (cf. Eq. \eqref{lin-approx-cond-nohs})
\begin{align}
\mathscr{D}&:= \frac{\tilde{R}^2 R  \alpha'}{\alpha \sinh \eta +  \alpha' R \cosh^2 \eta} = \frac{\varepsilon \tilde{R}^2}{\cosh \eta \left(\sinh \eta + \varepsilon \cosh \eta \right)},
\nn \\
\tilde{\mathscr{D}} &:= \mathscr{D} \left( \mathscr{D} \frac{\cosh^2 \eta}{\tilde{R}^2} -2 \right) \ .
\end{align}
Using Eq. \eqref{eff-metric-computation}, we can also derive the dilaton. To do this, we need the modulus $\vert \gamma \vert$ of the determinant of the metric $\gamma_{\mu \nu}$. By employing the Jacobi formula, we obtain
\begin{align}
\vert \gamma \vert  =   \frac{1}{\left(\alpha \sinh \eta\right)^8} \left(1+ \tilde{\mathscr{D}} \frac{\cosh^2 \eta}{\tilde{R}^2}\right),
\label{det-gamma-mu-nu}
\end{align}
in Cartesian coordinates,
which in view of Eqs.  \eqref{rho-M-G-relation} and \eqref{rho-M-cart} leads to
\begin{align}
\rho^2 &=   \alpha^4 \sinh^3 \eta \left(1+ \tilde{\mathscr{D}} \frac{\cosh^2 \eta}{\tilde{R}^2}\right)^{-1/2},
\label{rho-squared-exact}
\end{align}
  giving in the late-time regime the expansion
\begin{align}
\rho^2 \approx      \alpha^4 \cosh^3 \eta \left[1 + \varepsilon +  O(\varepsilon^3) \right]. 
\label{rho-squared-late-time}
\end{align}

It follows from Eqs. \eqref{frame-FLRW-hs-x} and \eqref{rho-squared-exact} that the inverse effective metric is
\begin{align}
  G^{\mu \nu}  &= \frac{1}{\rho^2} \left[\eta^{\mu \nu} \alpha^2  \sinh^2 \eta + t^\mu t^\nu\tilde{R}^2 R \alpha' \left(2 \alpha \sinh \eta +R \alpha'\cosh^2 \eta\right)   \right] 
  \nn \\
  &\approx  \frac{\alpha^2}{\rho^2} \left[\eta^{\mu \nu} \cosh^2 \eta + t^\mu t^\nu  (2 \varepsilon + \varepsilon^2) \tilde{R}^2\right] 
\nn \\
 & \approx  \frac{1}{\alpha^2 \cosh^3 \eta} \left[\eta^{\mu \nu} (1-\varepsilon + \varepsilon^2) \cosh^2 \eta + t^\mu t^\nu  (2 \varepsilon - \varepsilon^2)\tilde{R}^2 + O(\varepsilon^3)\right].
 \label{inverse-metric-hs}
\end{align}
We observe that both the frame and the metric are $\hs$ valued.
The following relation will also be useful in the following
\begin{align}
   \sqrt{\vert G \vert} \approx \left(\frac{\rho}{\alpha \cosh \eta}\right)^4 \left[1-\varepsilon+\varepsilon^2 + O(\varepsilon^3)\right],  
   \label{sqrt-det-G}
\end{align}
(in Cartesian coordinates),
which can be easily derived by means of Eqs. \eqref{eff-metric-computation} and  \eqref{det-gamma-mu-nu}. 

\subsection{Field strength and torsion}

In this section, we work in Cartesian coordinates $x^\mu$.
We first compute the field strength for the  background \eqref{FLRW-background-alpha}:
\begin{align}
   -\cF^{ab} = \{T^a,T^b\} &= 
    \{\a t^a,\a t^b\} \nn\\
 &= -\frac{\a^2}{\tilde R^2 R^2}\big[\theta^{ab}  
 + \tilde R^2 R \frac{\a'}{\a}(x^a t^b - t^a x^b)\big],
 \label{TT-CR-general}
\end{align}
which takes values in the spin 1 sector $\cC^1$ of $\hs$-valued functions on $\cM^{3,1}$.
This will be used below to evaluate the torsion tensor for late times $\eta \to \infty$. 

\subsubsection{Torsion on the unperturbed background}

The Weitzenb\"ock torsion of the unperturbed background with $\alpha = 1$ can be written as \cite{Steinacker:2020xph}
\begin{align}
     \tensor{\bar T}{_\r_\s^{\mu}} = \frac{1}{R^2\r^2}(\d^\mu_\s \t_\r - \d^\mu_\r \t_\s)
\end{align}
where $\t = \t^\mu\del_\mu$ is the cosmic timelike vector field, 
given by $\t^\mu = x^\mu$ in Cartesian coordinates, and $\tau_\mu = G_{\mu\nu}\t^\nu$.

The contraction of the torsion can be computed in general using 
\begin{align}
\tensor{T}{^{a}^{b}^{\mu}}\tensor{T}{_{a}_{b}^{\nu}} G_{\mu\nu}
   &=  \r^{2}  \tensor{T}{_\r_\s^{\mu}}\tensor{T}{_{\r'}_{\s'}^{\nu}} \g^{\r\r'} \g^{\s\s'} \g_{\mu\nu}
 %   = \tensor{T}{^\r_\s_{\mu}}\tensor{T}{_{\r}^{\s}_{\mu}} \g^{\mu\mu'}
    = \r^4 \tensor{T}{_{\r}^{\s}_{\mu}} \tensor{T}{^\r_\s_{\nu}} G^{\mu\nu},
    \label{torsion-contraction-frame-effective}
\end{align}
which for the unperturbed cosmic background gives 
\begin{align}
 \tensor{\bar T}{^\r_\s_{\mu}}\tensor{\bar T}{_{\r}^{\s}_{\nu}} G^{\mu\nu}&= \frac{6}{R^4\r^4} \t_\mu \t^\mu 
  \approx -\frac{6}{R^2\r^2}. 
    \label{torsion-contraction-frame-G}
\end{align}
As a check, we also find using the frame formalism 
\begin{align} \label{bar-torsion-1}
\bar{T}^{ab \mu} &= \frac{1}{R^2} (\eta^{a\mu} x^b - \eta^{b\mu} x^a)   %\approx  \frac{\bar a^4}{R^6\cosh^6\eta} (\eta^{a\mu} x^b - \eta^{b\mu} x^a),
\end{align}
which gives
\begin{align}
    \tensor{\bar T}{^{a}^{b}^{\mu}}\tensor{\bar T}{_{a}_{b}^{\nu}} G_{\mu\nu}
   &= \frac{1}{R^4} 
   (\eta^{a\mu} x^b - \eta^{b\mu} x^a)
   (\eta^{a\nu} x^b - \eta^{b\nu} x^a) G_{\mu\nu} \nn\\
    &= -\frac{6}{R^4} R^2 \cosh^2(\eta) \sinh(\eta)
 \approx  -\frac{6}{R^2} \r^2 \ ,
\end{align}
which is consistent with Eq. \eqref{torsion-contraction-frame-G} due to Eq. \eqref{torsion-contraction-frame-effective}.

\subsubsection{$\hs$-valued torsion on the deformed background $T^a = \a t^a$}

The torsion of the background \eqref{FLRW-background-alpha}   is obtained using Eqs.  \eqref{torsion-F-identity} and \eqref{TT-CR-general} as 
\begin{align}
T^{ab \mu} &= \{\cF^{ab},x^\mu\} = \a^2\bar{T}^{ab \mu} \left(1+\frac{\alpha'}{\alpha}x^4 \right) -\frac{2 \alpha \alpha'}{R} \theta^{ab} t^\mu 
 \nn \\
&- \frac{\alpha' \alpha}{R} \left(\theta^{a \mu} t^b - \theta^{b \mu} t^a \right) -\tilde{R}^2 t^\mu \left(x^a t^b  - x^b t^a\right) \left(\alpha \alpha'' + \alpha'^{ \,2} \right) \ .
\label{torsion-1}
\end{align}
Using  the  relation \cite{Steinacker:2020xph} 
\begin{align}
\theta^{\mu \nu}  \approx \frac{\tilde{R}^2}{\cosh \eta} \left(x^\mu t^\nu-x^\nu t^\mu\right),  \quad \eta \to + \infty
\end{align}
valid in the late-time regime, Eq. \eqref{torsion-1} yields 
\begin{align}
T^{ab \mu} &\approx \a^2\bar{T}^{ab \mu}(1+\varepsilon )
    - \tilde{R}^2 \alpha^2 \left[\frac{\alpha''}{\alpha} + \Big(\frac{\alpha'}{\alpha}\Big)^{\, 2} +3 \frac{\alpha'}{\alpha}\frac{1}{R \cosh \eta}\right]  t^\mu \left(x^a t^b-x^b t^a \right)  \nn\\
    & \approx \a^2\bar{T}^{ab \mu}(1+\varepsilon )
    - \tilde{R}^2 \frac{\bar a^4}{R^6 \cosh^8 \eta} \left(\frac{\alpha''}{\alpha} R^2 \cosh^2 \eta + \varepsilon^{2} + 3 \varepsilon\right)  t^\mu \left(x^a t^b-x^b t^a \right) \nn\\
    &=: \a^2 \bar{T}^{ab \mu}(1+\varepsilon )
    + T^{ab \mu}_{(2)},
    \label{torsion-hs}
\end{align}
where we have exploited  Eqs.   \eqref{lin-approx-cond-nohs} and \eqref{bar-a}, and denoted the higher-spin components of the torsion with $ T^{ab \mu}_{(2)}$. 
We can estimate their size in the late-time regime as follows:
\begin{align}
    T^{ab \mu}_{(2)} &= O\left(\frac{\bar a^4}{R^5\cosh^5 \eta } \ \varepsilon\right),  \nn\\
   \a^2\bar{T}^{ab \mu} &=  O\left( \frac{\bar a^4}{R^5\cosh^5\eta}(1 + \varepsilon)\right).
\end{align}
Notice  that here we consider $\frac{\alpha''}{\alpha}$ to be of the same order as $\big(\frac{\alpha'}{\alpha}\big)^{\, 2}$.

At this stage, we are ready to evaluate the term $\tensor{T}{_\r^\s_{\mu}}\tensor{T}{^\r_\s_{\nu}} G^{\mu\nu}$ which will occur in the one-loop effective action (see Eq. \eqref{ind-grav-disc-1}, below). To this end, let us first calculate $ T^{ab\mu} \tensor{T}{_a_b^\nu}$. Bearing in mind Eqs. \eqref{bar-torsion-1} and \eqref{torsion-hs},  we obtain after a lengthy calculation 
\begin{align}
 T^{ab\mu} \tensor{T}{_a_b^\nu} &=   -\frac{2 \alpha^4}{R^4}(1+\varepsilon)^2 \left(\eta^{\mu \nu} R^2 \cosh^2 \eta + x^\mu x^\nu\right) 
 \nn \\
 &+2\mathscr{C} \cosh^2 \eta \left[2 \alpha^2 (1+\varepsilon) - \mathscr{C} \frac{R^2 \cosh^2\eta}{\tilde{R}^2}\right] t^\mu t^\nu,
 \label{contraction-1}
\end{align}
%\hcs{this is exact}
where we have employed Eq. \eqref{relations-Cartesian} and we have defined 
\begin{align}
\mathscr{C}:=   - \frac{\alpha^2 \tilde{R}^2}{R^2 \cosh^2 \eta} \left(\frac{\alpha''}{\alpha}R^2 \cosh^2 \eta + \varepsilon^2 +3 \varepsilon\right). 
\end{align}
Owing to Eqs.  \eqref{eff-metric-computation} and \eqref{rho-squared-late-time}, Eq. \eqref{contraction-1} leads to 
\begin{align}
 T^{ab\mu} \tensor{T}{_a_b^\nu} G_{\mu \nu} \approx  \frac{2 \alpha^6}{R^2 } \cosh^3 \eta \left[-3 -13  \varepsilon -17 \varepsilon^2 - 2 \frac{\alpha''}{\alpha} R^2 \cosh^2 \eta + O(\varepsilon^3)\right]
\end{align}
for late times, 
and together with Eq. \eqref{torsion-contraction-frame-effective} the desired term $\tensor{T}{_\r^\s_{\mu}}\tensor{T}{^\r_\s_{\nu}} G^{\mu\nu}$ is obtained as 
\begin{align}
 \tensor{T}{_\r^\s_{\mu}}\tensor{T}{^\r_\s_{\nu}} G^{\mu\nu} &\approx   \frac{2}{\alpha^2 R^2 \cosh^3 \eta} \left[-3-7 \varepsilon - 2 \frac{\alpha''}{\alpha} R^2 \cosh^2 \eta + O(\varepsilon^3)\right]
 \label{torsion-contraction-final-pert}
\end{align}
in the late-time regime.
As a consistency check we reconsider the ansatz without $\hs$ components, where the divergence constraint \eqref{div-constraint-E} led to Eq.  \eqref{A-rhoM-div}, which in turn implies that $\a = const$. Therefore, it follows from Eqs. \eqref{eff-metric-and-A} and  \eqref{torsion-expr} that for late times
\begin{align}
 \tensor{T}{_\r^\s_{\mu}}\tensor{T}{^\r_\s_{\nu}} G^{\mu\nu} &\approx -\frac{6}{\a^2 R^2 \cosh^3 \eta} \ .
 \label{torsion-contraction-final-unperturbed}
\end{align}
This is consistent both with the unperturbed background calculation  \eqref{torsion-contraction-frame-G}, and with the leading-order term of Eq. \eqref{torsion-contraction-final-pert}.

\section{Evaluation of the classical  action}

From now on, 
we consider a background brane with product structure 
\begin{align}
\cM^{3,1} \times \cK \ \subset \R^{9,1}
\label{fuzzy M-K-BG}
\end{align}
embedded in flat (uncompactified\footnote{All propagating modes are confined to the brane in the matrix model at weak coupling. Therefore there is no need to compactify target space.}) target space, as required for the present mechanism for gravity.
Here $\cM^{3,1}$ plays the role of spacetime
 given by the generalized FLRW background \eqref{FLRW-background-alpha}
\begin{align}
T^a = \a(\eta) \, t^a
\label{FLRW-background-alpha-second-time}    
\end{align}
 and $\cK = \cK_N$ are fuzzy extra dimensions which support only  finitely many degrees of freedom, see section \ref{sec:fuzzy-K}.
This means that the total Hilbert space is given by $\cH_\cM \otimes \C^N$, where $\C^N$ is the Hilbert space for the extra dimensions with $N \in \N$.
The $\cK$ factor is essential to obtain an induced Einstein-Hilbert term in the one-loop effective action, with effective Newton constant set by the Kaluza-Klein scale of $\cK$.

On such a background, the Yang-Mills (YM) action can be written in the semiclassical regime as %\be{added the index $i$ in the equation}
\begin{align}
 S_{\rm YM} = - \int\limits_\cM \Omega\, \tr_\cK(\cF_{a b}
 \cF^{a b} +  \cF_{ij}\cF^{ij} + 2 \{T^{a},T^i\} \{T_{a},T_i\}) \ \qquad i=4,\ldots,9\ 
 \label{YM-action-semiclass}
\end{align}
dropping the fermions for now. Here   the invariant (symplectic) volume form $\Omega = d^4 x \, \rho^{-2} \sqrt{\vert G \vert } $ on $\cM^{3,1}$ can be written as  
\begin{align}
 \Omega   &=  \frac{1}{\sinh\eta}   d x^0 \dots d x^3
 =  \cosh^3\eta d\eta  \sinh^2  \chi d\chi   \sin \theta d \theta d\varphi
  \label{measure-H4}
 \end{align}
 in hyperbolic coordinates.
The first term in Eq. \eqref{YM-action-semiclass} is the bare action for the cosmological background $\cM^{3,1}$, the second  represents the contribution from the extra dimensions $\cK$, and the last \qm{mixing} term denotes a kinetic term for time-dependent $\cK$. We will evaluate these contributions separately in the following.

\subsection{Contributions from the FLRW spacetime $\cM^{3,1}$}

In this section, we work in Cartesian coordinates $x^\mu$.
Using the field strength \eqref{TT-CR-general}, we can obtain the YM term for the background \eqref{FLRW-background-alpha-second-time} as
\begin{align}
   \cF^{ab}  \cF_{ab} &= \frac{\a^4}{\tilde R^4 R^4} \big[\theta^{ab}  
 + \tilde R^2 R \frac{\a'}{\a}(x^a t^b - t^a x^b)\big]\big[\theta_{ab}  
 + \tilde R^2 R \frac{\a'}{\a}(x_a t_b - t_a x_b)\big] \nn\\
%  &= \frac{\a^4}{\tilde R^4 R^4} \big(\theta^{ab}\theta_{ab} - 4 \tilde R^2 R^3\frac{\a'}{\a} \sinh(\eta)\cosh^2(\eta)   +2 \tilde R^4 R^2\frac{\a'^2}{\a^2}(xx)(tt)   \big) \nn\\
%  &= \frac{2\a^4}{\tilde R^2 R^2} \big(2 - \cosh^2(\eta) - 2 R\frac{\a'}{\a} \sinh(\eta)\cosh^2(\eta)    - R^2\frac{\a'^2}{\a^2} \cosh^4(\eta)   \big) \nn\\
%    &\approx \frac{2\a^4}{\tilde R^2 R^2} \big(2 - \cosh^2(\eta)  - 2 \varepsilon\cosh^2(\eta)    - \varepsilon^2 \cosh^2(\eta)   \big) \nn\\
  &\approx \frac{2\a^4}{\tilde R^2 R^2} \big[2 
   - (1 + \varepsilon)^2 \cosh^2\eta \big] 
   \label{F-ab-squared}
\end{align}
%%V3  sign in front of $\varepsilon$ changed
in the late time regime,
using
\begin{align}
\theta^{ab} x_a t_b &=  x_a x^a \sinh\eta= - R^2 \sinh\eta \cosh^2\eta, 
\nn\\
\theta^{ab}\theta_{ab} &= 4\tilde R^2 R^2 - \tilde R^4 R^2 t^a t_a + \tilde R^2 x^a x_a  = 2\tilde R^2 R^2 (2 - \cosh^2\eta), 
\end{align}
along with Eq. \eqref{relations-Cartesian}. 
The corresponding contribution to the YM action can be written as 
\begin{align}
 - \int \Omega\, \tr_\cK(\cF^{ab}\cF_{ab}) 
 &\approx -  \frac{2N}{\tilde R^2 R^2} \int \Omega\,  \a^4 \left[2 
- (1 + \varepsilon)^2 \cosh^2\eta\right] 
\nn \\
&= -  \frac{2N V_3}{\tilde R^2 R^2} \int d \eta \, \a^4 \cosh^3 \eta  \left[2 
 - \left(1 + \frac{1}{\alpha} \frac{d \alpha}{ d \eta}\right)^2 \cosh^2\eta\right],
\end{align}
%%V3: sign in last line changed
where
\begin{align}
 V_3 := \int   \sinh^2  \chi  \sin \theta d \chi    d \theta d \varphi
\end{align}
is a spacelike volume factor,
and $N$ arises from the trace over the extra dimensions. 
This geometric YM contribution can be interpreted as brane tension of $\cM^{3,1}$, which will be seen to be dominant at late times. 
Remarkably, this large tension is not in conflict with obtaining massless spin 2 excitations on the brane \cite{Sperling:2019xar}.

\subsection{Contributions from fuzzy extra dimensions $\cK$}
\label{sec:fuzzy-K}

We recall that the IKKT model comprises 10 matrices $T^A$. Among these, the first 3+1 are used to describe space-time. 
The fuzzy extra dimensions are realized by the remaining 6 transversal matrices $T^i$, which acquire a nontrivial vacuum expectation value interpreted as (fuzzy) embedding function
\begin{equation}\label{II-Ti}
    T^i = m_\cK \cK^i \sim z^i: \quad \cK \hookrightarrow \R^6\ ,\qquad i=4,\ldots,9\ .
\end{equation}
%%V3: def of $\cK^i$ inserted
The precise structure of $\cK$ is not relevant for our discussion except for the finite, discrete (positive) spectrum, labelled by $\Lambda$, of its Laplacian $\Box_{\cK}$ arising form the splitting $\Box = \Box_{\cM^{1,3}}+\Box_{\cK}$:
\begin{equation}\label{KKmass}
    \Box_{\cK}\,\lambda_{\Lambda} \ = m_{\Lambda}^2\,\lambda_{\Lambda}\ ,\quad m_{\Lambda}^2 \ = m_{\cK}^2\mu_{\Lambda}^2
\end{equation}
associated with eigenmodes $\lambda_{\Lambda}\in \End(\cH_{\cK})$. Here $\mu_\Lambda$ is dimensionless, while $m_\cK^2$ sets the scale of $\cK$.
Then the second term in the YM action \eqref{YM-action-semiclass} contributes to the potential for $m_\cK^2$  via
\begin{align}
\tr_\cK(\cF_{ij}\cF^{ij}) &=: m_\cK^4 F_\cK^2 \ , 
\end{align}
where $F_\cK^2$ is a  discrete number depending on the structure of $\cK$.
Therefore
\begin{align}
     \int \limits_{\mathcal{M}} \Omega \ 
 \tr_\cK(\cF_{ij} \cF^{ij})
    &=  \int \limits_{\mathcal{M}} \Omega \ m_\cK^4 F_\cK^2 \ .
    \label{YM-K-term}
\end{align}

\subsection{Mixed term or kinetic term for $m_\cK$ }

The last (mixed) contribution in  Eq. \eqref{YM-action-semiclass} amounts to a kinetic term for $m_\mathcal{K}$, which can be evaluated as
\begin{align}
-2 \int \limits_{\mathcal{M}} \Omega\, \tr_\cK \{T^\alpha,T^i\}  \{T_\alpha,T_i\} = -  2 \int \limits_{\mathcal{M}} d^4 x \sqrt{\vert G \vert } \, d_\cK\,  \partial^\mu m_\mathcal{K} \partial_\mu m_\mathcal{K} \ 
\label{kinetic-fuzzy}
\end{align}
where
\begin{align}
 d_\cK := \tr_\cK(\cK^i \cK_i) \ ,
\end{align}
is a discrete number depending on the structure of $\cK$. 
Let us evaluate this term explicitly for late times. First of all, by using Eq.  \eqref{inverse-metric-hs}   we find, up to corrections $O(\varepsilon^3)$, 
\begin{align}
 G^{\mu \nu} \partial_\mu m_\cK \partial_\nu m_\cK   \approx \rho^{-2} \alpha^2 \left[ \eta^{\mu \nu} \cosh^2 \eta + t^\mu t^\nu \tilde{R}^2 \left(2 \varepsilon + \varepsilon^2 \right)\right] \partial_\mu m_\cK \partial_\nu m_\cK,  
\end{align}
which upon employing Eq. \eq{time-derivative-cartes} jointly with the identities \eqref{x-mu-x-mu-prod} and \eqref{x-mu-t-mu} yields
\begin{align}
G^{\mu \nu} \partial_\mu m_\mathcal{K} \partial_\nu m_\mathcal{K} \approx -\frac{\alpha^2}{\rho^2 R^2}  \left(\frac{d m_{\cK}(\eta)}{d \eta}\right)^2 \ . 
\label{G-mu-nu-del-mk}
\end{align}
Therefore, using Eqs. \eqref{rho-squared-late-time} and \eqref{sqrt-det-G} and neglecting $O(\varepsilon^3)$ corrections we finally obtain 
\begin{align}
-   \int \limits_{\mathcal{M}} d^4 x \sqrt{\vert G \vert } \, d_\cK\,  \partial^\mu m_\mathcal{K} \partial_\mu m_\mathcal{K} \approx \frac{d_\cK V_3}{ R^2} \int d \eta \, \alpha^2 \cosh^3 \eta \left[1+O(\varepsilon^3)\right] \left(\frac{d m_{\cK}(\eta)}{d \eta}\right)^2. 
\end{align}

Notice that Eq. \eqref{G-mu-nu-del-mk} is consistent  with an evaluation using  the classical metric
without the $\hs$ components, which for $SO(3,1)$-invariant frames gives
\begin{align}
G^{\mu \nu} \partial_\mu m_\mathcal{K} \partial_\nu m_\mathcal{K} = -A \left(\frac{1}{R \sinh \eta} \frac{d m_{\mathcal{K}}(\eta)}{d \eta}\right)^2,
\label{kinetic-term-no-hs}
\end{align}
exploiting Eq. \eqref{eff-metric-FRW}.
That term clearly suppresses any variations of $m_\cK$.

%For $E(3)$-invariant frames we obtain
%\begin{align}
%G^{\mu \nu}     \partial_\mu m_\mathcal{K} \partial_\nu m_\mathcal{K}= -B^3 \left(\frac{\dot{m}_{\mathcal{K}}(t)}{A}\right)^2
%\end{align}
%as a trivial consequence of Eq. \eqref{effective-metric-E-3-inv} (recall that dot is derivative with respect to $x^0=t$). 

\paragraph{Scalar matter contribution.}

Even though we do not consider matter in this paper, it is not hard to see that matter will typically not significantly affect the late-time cosmic evolution, in contrast to general relativity. To see this, it suffices to consider the contribution from nonabelian scalar fields, which would have an action similarly as in Eq. \eqref{kinetic-fuzzy}, with $T^i \sim \phi$ viewed as (nonabelian) scalar fields.
Then their contribution to the action 
is much smaller than the contribution of the 
YM brane tension of $\cM^{3,1}$
\begin{align}
    \{T^\alpha,T^i\} \{T_\alpha,T_i\} \sim \rho^2 G^{\mu\nu} \partial_\mu \phi \partial_\nu \phi 
    \ \ll \  \{T^\alpha,T^\beta\} \{T_\alpha,T_\beta\} \sim  \a^4\cosh^2\eta \ 
\end{align}
for large $\eta$;
note that the energy density $\sim G^{\mu\nu} \partial_\mu \phi \partial_\nu \phi $  of matter  will decay like
$\frac 1{a(t)^3} \sim \frac 1{\rho^3}$ with the cosmic expansion, assuming $\a = 1$. The inclusion of quantum corrections through $\alpha(\eta)$ will not change this conclusion.

\subsection{Combined classical YM action}

Summing up all the above terms,   the YM action \eqref{YM-action-semiclass} becomes 
\begin{align}
S_{\rm YM} &\approx V_3 \Biggl\{-F^2_{\cK} \int d \eta \, m^4_{\cK} \cosh^3 \eta +\frac{d_\cK }{ R^2} \int d \eta \, \alpha^2 \cosh^3 \eta \left[1+O(\varepsilon^3)\right] \left(\frac{d m_{\cK}}{d \eta}\right)^2  
\nn \\ 
&-\frac{2N }{\tilde R^2 R^2} \int d \eta \, \a^4 \cosh^3 \eta  \left[2 - \left(1 + \frac{1}{\alpha} \frac{d \alpha}{ d \eta}\right)^2 \cosh^2\eta\right]\Biggr\}
\label{YM-action-eta}
\end{align}
 in the late-time regime.
%%V3: sign in last line changed

\section{One-loop effective action}

We want to understand the dynamical evolution of the above FLRW spacetime \eqref{FLRW-background-alpha},
%\be{let's refer to an equation}, 
realized as brane solution of the IKKT model, taking into account quantum effects at one loop. The one-loop effective action
on covariant spacetime branes was computed in Ref. \cite{Steinacker:2023myp}. As pointed out before, we need to assume that the background has a product structure $\cM^{3,1} \times \cK$ as in Eq. \eqref{fuzzy M-K-BG} to obtain the Einstein-Hilbert action at one loop.
Then the combined geometric one-loop effective action
on the spacetime brane $\cM^{3,1}$  takes the form
\begin{align}
 S_{\rm 1loop} = S_{\rm YM} + S_{\rm grav} + S_{\rm vac} \ .
 \label{S-eff-1loop}
\end{align}
Here
\begin{align}
 S_{\rm YM} = -  \frac{1}{g^2}\int\limits_\cM \Omega\, \tr_\cK\cF_{AB}\cF^{AB}
\end{align}
is the (bosonic part of the) bare action of the IKKT model with \qm{field strength}
\begin{align}
    \cF^{AB} = i [T^A,T^B] \sim -\{T^A,T^B\}
\end{align}
(which includes contributions from $\cK$ computed above),
while 
the induced gravitational action at one loop has the form 
\begin{align}
 S_{\rm grav}
  &= -\frac{1  }{(2\pi)^4}\int\limits_\cM d^4 x \sqrt{\vert G\vert }\,\r^{-2} m_\cK^2 c^2_{\cK} \,
  \tensor{T}{_\r^\s_{\mu}}\tensor{T}{^\r_\s_{\nu}} G^{\mu\nu} \
  +   \Gamma_{\rm h.o.}(\cM,\cK) \ ,
   \label{ind-grav-disc-1}
\end{align}
where the second term indicates higher-order contributions from the one-loop effective action.
This determines the effective Newton constant as \cite{Steinacker:2023myp}
\begin{equation}\label{GNdef}
    G_N = \frac{\pi^3\rho^2}{2c_\cK^2m_\cK^2} \ ,
\end{equation}
where $c^2_{\cK}$ is a (large) constant depending on the structure of $\cK$.
Note that the coupling to matter
-- including  fermions\footnote{Fermions do indeed couple 
appropriately to the background geometry, as discussed in Refs. \cite{Battista:2022vvl,Battista:2023kcd}.}, but
also  the nonabelian bosonic sector 
arising from $\cK$ 
-- is contained in $S_{\rm YM}$.
Finally, 
\begin{align}
   S_{\rm vac} =  \Gamma_{\rm 1 loop}^\cK
    + \Gamma_{\rm 1 loop}^{S^2}
    +  \Gamma_{\rm 1 loop}^{\cK - S^2}
    \label{S-vac}
\end{align}
subsumes the remaining contributions of the induced vacuum energy
which are independent of the geometry of $\cM^{3,1}$, where 
\begin{subequations}
\begin{align}
\Gamma_{\rm 1 loop}^\cK
 &\approx -\frac{\pi^2 }{2(2\pi)^4}\,
  \int\limits_\cM \Omega\, \r^{-2} m_\cK^4
  C_1   \  ,
   \label{1-loop-pot-cK-cov}  \\
 \Gamma_{\rm 1 loop}^{S^2}
 &\approx -\frac{\pi^2 }{2(2\pi)^4}
  \int\limits_\cM \Omega\, \r^{-2} \frac {C_2}{m^4_\cK R^8} 
   \
  \label{tr-S2-cov-simp} \ , \\
   \Gamma_{\rm 1 loop}^{\cK - S^2}
  %= \frac{3i}4\Tr\Big(\frac{V_{4,\cK}}{(\Box -i \varepsilon)^4}\Big)
  &\approx -\frac{1}{32\pi^2}
  \int\limits_\cM \Omega \, \r^{-2}\frac{C_3}{R^4}
   \ ,
   \label{tr-S2-K-mix-simple}
\end{align}
\end{subequations}
are  the contributions
from $\cK$, from $S^2_n$, 
 and  mixed contributions
from $S^2_n - \cK$, respectively.
%Here  we  introduced the following definitions: 
%\begin{align}
%C_1 &:= \frac{1}{4} \sum_{\L;s} \frac{(2s+1) V_{4,\L}}{\mu^4_{\L}}\ , \\
%C_2 &:= \frac{1}{4 } \sum\limits_{\L;s} \frac{\tilde V_{4,s}}{\mu^4_{\L}} \ , 
%\\
%C_3 &:= \sum_{\L;s}   \frac{s(s+2)(2s+1) C^2_{\L}}{\mu^4_{\L}} \ .
%\end{align}
%Here  $V_{4,\L}$, $ \tilde V_{4,s}$ and $C^2_{\L}$ 
Here $C_1, C_2$ and $C_3$
are (large) constants determined by the structure of $\cK$ \cite{Steinacker:2023myp}.

\paragraph{Gravitational action without $\hs$ components.}

We first evaluate the above gravitational action using the geometrical results in section \ref{sec:so31-frames}, assuming that the $\hs$ components of the frame and torsion can be neglected. However, this is justified
only in the linearized regime, as long as the $\hs$ components are negligible. 
Bearing in mind Eqs. \eqref{constraint-H4}, \eqref{eff-metric-and-A}, and  \eqref{torsion-expr}, the gravitational action \eqref{ind-grav-disc-1} then becomes 
\begin{align}
 S_{\rm grav}
&= \frac{6 c^2_{\cK} }{(2\pi)^4 R^2 }\int\limits_\cM d^4 x \sqrt{\vert G\vert }\,\r^{-2} m_\cK^2  \,
A^{-1} \left(\frac{\partial_\eta A}{\sinh \eta}\right)^2 \ +   \Gamma_{\rm h.o.}(\cM,\cK)  \nn\\
& \approx \frac{6 c^2_{\cK} }{(2\pi)^4 R^2 }\int\limits_\cM d^4 x \, m_\cK^2 \left(\frac{1}{\alpha \cosh^2 \eta}\right)^2 \ +   \Gamma_{\rm h.o.}(\cM,\cK),
\label{S-grav-lorentz-inv}
\end{align}
where in the last line, which is valid for $\eta \to + \infty$,  we have used Eqs. \eqref{rho-equation} and \eqref{A-expression} assuming that $\alpha$ is constant.
% whereas for $E(3)$-invariant frames we have
%\begin{align}
% S_{\rm grav}
%&= \frac{6 c^2_{\cK} }{(2\pi)^4 }\int\limits_\cM d^4 x \sqrt{\vert G\vert }\,\r^{-2} m_\cK^2  \,
%B \left(\frac{\dot B}{A}\right)^2 \ +   \Gamma_{\rm h.o.}(\cM,\cK), (...)
%\label{S-grav-E-3-inv}
%\end{align}
%upon exploiting Eqs. \eqref{effective-metric-E-3-inv} and \eqref{torsion-E-3-inv}. 
This will be refined below by taking into account the $\hs$ contributions.

\section{FLRW background and solution at one-loop level}
\label{sec:FLRW-one-loop}

In this section, we will evaluate the one-loop effective action  \eqref{S-eff-1loop}
for the deformed FLRW background \eqref{FLRW-background-alpha}, taking into account also the $\hs$ contributions.  This will be used to derive the equations of motion for the cosmic scale parameter $a(t)$.

Let us analyze the various terms occurring in the one-loop effective action $S_{\rm 1loop}$ separately. Due to the presence of higher spin components, it is safer to use Eq. \eqref{ind-grav-disc-1} for the one-loop gravitational action, rather than its Einstein-Hilbert form. This can be evaluated using  Eqs. \eqref{rho-squared-late-time},  \eqref{sqrt-det-G}, and  \eqref{torsion-contraction-final-pert}, which give 
\begin{align}
 S_{\rm grav}
  &\approx \frac{6 c_{\mathcal{K}}^2 }{(2\pi)^4 R^2}\int\limits_\cM d^4 x \left(\frac{m_\cK }{\alpha \cosh^2 \eta}\right)^2 \left[1 + \frac{7}{3} \varepsilon +\frac{2}{3} \frac{\alpha''}{\alpha} R^2 \cosh^2 \eta + O(\varepsilon^3)\right]
  +   \Gamma_{\rm h.o.}(\cM,\cK)
  \label{S-grav-2}
\end{align}
in the late-time regime. Transforming the derivatives with respect to $x^4$ to derivatives with respect to $\eta$ (cf. Eq. \eqref{x4-coord}) and using Eq. \eqref{measure-H4},  $S_{\rm grav}$ can be written as 
\begin{align}
 S_{\rm grav}
  &\approx \frac{6 c_{\mathcal{K}}^2  V_3}{(2\pi)^4 R^2}\int d \eta \left(\frac{m_\cK }{\alpha }\right)^2 \left[1+\frac{5}{3 \alpha} \frac{d \alpha}{d \eta}  +\frac{2}{3\alpha} \frac{d^2 \alpha}{d \eta^2} + O(\varepsilon^3)\right]
  +   \Gamma_{\rm h.o.}(\cM,\cK),
  \label{S-grav-eta-funct}
\end{align}
which  after partial integration, assumes the first-order form
\begin{align}
 S_{\rm grav}
  &\approx \frac{6 V_3 c_{\mathcal{K}}^2}{(2\pi)^4 R^2}\int d \eta \left(\frac{m_\cK }{\alpha }\right)^2 \left[1+\frac{5}{3 \alpha} \frac{d \alpha}{d \eta}   + O(\varepsilon^3)\right] 
  \nn \\
 & - \frac{4  V_3 c_{\mathcal{K}}^2}{(2\pi)^4 R^2}\int d \eta \, \frac{d}{d \eta} \left(\frac{m^2_{\mathcal{K}}}{\alpha^3}\right) \frac{d \alpha}{d \eta}
  +   \Gamma_{\rm h.o.}(\cM,\cK) \ .
 \label{S-grav-first-order-eta} 
\end{align}
For the YM piece, we employ the calculation performed before, see Eq. \eqref{YM-action-eta}.  Finally, the one-loop vacuum term \eqref{S-vac} gives
\begin{subequations}
\label{S-vac-eta}
\begin{align} 
 \Gamma_{\rm 1 loop}^\cK
&\approx -\frac{\pi^2 V_3}{2(2\pi)^4}\,
  \int d \eta \frac{m_\cK^4}{\alpha^4} \left[1-\frac{1}{\alpha} \frac{d \a}{d \eta}+ \left(\frac{1}{\alpha} \frac{d \a}{d \eta}\right)^2 + O(\varepsilon^3)\right] C_1  \  ,
  \\
\Gamma_{\rm 1 loop}^{S^2}
 &\approx -\frac{\pi^2 V_3 }{2(2\pi)^4 R^8}
  \int d \eta  \frac{1}{ \alpha^4 m^4_\cK } \left[1-\frac{1}{\alpha} \frac{d \a}{d \eta}+ \left(\frac{1}{\alpha} \frac{d \a}{d \eta}\right)^2 + O(\varepsilon^3)\right]  C_2 \
 \ , \\
   \Gamma_{\rm 1 loop}^{\cK - S^2}
 &\approx -\frac{\pi^2V_3}{2(2\pi)^4 R^4}
  \int d \eta \frac{1}{\alpha^4}\left[1-\frac{1}{\alpha} \frac{d \a}{d \eta}+ \left(\frac{1}{\alpha} \frac{d \a}{d \eta}\right)^2 + O(\varepsilon^3)\right]
 C_3 \ , 
\end{align}
\end{subequations}
using Eqs. \eqref{measure-H4} and \eqref{rho-squared-exact}.

\subsection{Lagrangian and equations of motion}
\label{sec:eom-full}

In view of Eqs. \eqref{YM-action-eta}, \eqref{S-grav-first-order-eta}, and \eqref{S-vac-eta}, the action 
$S= \int d\eta \cL$ can be written  in the late-time regime in terms of an effective Lagrangian
having the form
\begin{align}
\mathcal{L} &= \mathcal{L}_{\rm 1loop} \left(q^{\underline{a}},\frac{d q^{\underline{a}}}{d \eta}\right) + \mathcal{L}_m \left( \psi,\frac{d \psi}{d \eta}, q^{\underline{a}},\frac{d q^{\underline{a}}}{d \eta}\right), 
\end{align}
where (hereafter we will omit the $O(\varepsilon^3)$ symbol) 
\begin{align}
\mathcal{L}_{\rm 1loop} &= \mathcal{L}_{\rm grav}  + \mathcal{L}_{\rm YM} + \mathcal{L}_{\rm vac},
\label{Lagrangian-grav-YM-vac}
\\
 \mathcal{L}_{\rm grav} & \approx  \frac{2 V_3 c_{\mathcal{K}}^2}{(2\pi)^4 R^2} \left(\frac{m_\cK }{\alpha }\right)^2 \Biggl[  3+\frac{5}{ \alpha} \frac{d \alpha}{d \eta}   +6 \left(\frac{1}{\alpha} \frac{d \alpha }{d \eta} \right)^2
 -  4 \left( \frac{1}{\alpha} \frac{d \alpha}{d \eta} \right) \left( \frac{1}{m_\cK} \frac{d m_\cK}{d \eta}\right)\Biggr] + \Gamma_{\rm h.o.}(\cM,\cK),
 \label{Lagr-grav-1}
 \\
 \mathcal{L}_{\rm YM} &\approx  \frac{V_3}{R^2} \alpha^2 \cosh^3 \eta \Biggl\{-F^2_{\cK} R^2  \frac{m^4_{\cK}}{\alpha^2}  +d_\cK  \left(\frac{d m_{\cK}}{d \eta}\right)^2  - \frac{2N }{\tilde R^2 } \a^2   \left[2 - \left(1 + \frac{1}{\alpha} \frac{d \alpha}{ d \eta}\right)^2 \cosh^2\eta\right]\Biggr\},
 \\
 \mathcal{L}_{\rm vac}   &\approx -\frac{\pi^2 V_3}{2(2\pi)^4} \frac{1}{\alpha^4}\left[1-\frac{1}{\alpha} \frac{d \a}{d \eta}+ \left(\frac{1}{\alpha} \frac{d \a}{d \eta}\right)^2 \right] \Biggl( C_1 m_\cK^4 + \frac{C_2}{m_\cK^4 R^8 } + \frac{C_3}{R^4} \Biggr) \ .
\label{L-vac-1}
\end{align}
Here $\mathcal{L}_m$ is the matter Lagrangian for some generic matter field $\psi$,  and  we have adopted the compact notation
\begin{align}
  q^{\underline{a}}:= \{ \alpha(\eta), m_{\mathcal{K}}(\eta) \} \ , \qquad \underline{a} =1,2
\end{align}
for the geometric degrees of freedom of interest.
 Then the Euler-Lagrange equations 
\begin{align}
\frac{d}{ d \eta} \frac{\partial \mathcal{L}}{\partial(\partial_\eta q^{\underline{a}})} -\frac{\partial \mathcal{L}}{\partial q^{\underline{a}}}=0
\end{align}
yield -- after a lengthy calculation -- the following equations of motion for $\a$ and $m_\cK$: 
\begin{align}
&\frac{2 c^2_\cK}{(2 \pi)^4 R^2} \frac{m_\cK^2}{\alpha^3} \Biggl[ \frac{24}{m_\cK \alpha} \frac{d \alpha}{ d\eta} \frac{d m_\cK}{ d\eta} -\frac{24}{\alpha^2} \left(\frac{d \alpha}{ d\eta}\right)^2 - \frac{4}{m_\cK^2} \left(\frac{d m_\cK}{ d\eta}\right)^2 + \frac{10}{m_\cK} \frac{d m_\cK}{ d\eta} + \frac{12}{\alpha} \frac{d^2 \alpha}{ d\eta^2} - \frac{4}{m_\cK} \frac{d^2 m_\cK}{ d\eta^2} 
\nn \\
&+ 6 \Biggr] - \frac{4N}{R^2 \tilde{R}^2} \alpha^3 \cosh^5 \eta \Biggl[-3-\frac{5}{\alpha} \frac{d \alpha}{ d\eta} -\frac{1}{\alpha^2} \left(\frac{d \alpha}{ d\eta}\right)^2 -\frac{1}{\alpha} \frac{d^2 \alpha}{ d\eta^2} + \frac{d_\cK \tilde{R}^2}{2N} \frac{1}{\alpha^2 \cosh^2 \eta} \left(\frac{d m_\cK}{d \eta}\right)^2 
\nn \\
&- \frac{4}{\cosh^2 \eta}\Biggr]  -\frac{\pi^2 }{2(2\pi)^4} \frac{1}{\alpha^5}\Biggl\{\Biggl[4-\frac{6}{\alpha^2} \left(\frac{d \alpha}{ d\eta}\right)^2 +\frac{2}{\alpha} \frac{d^2 \alpha}{ d\eta^2}\Biggr]
 \Biggl(  C_1 m_\cK^4 + \frac{C_2}{m_\cK^4 R^8 } + \frac{C_3}{R^4} \Biggr) 
\nn \\ 
&+ \left(\frac{2}{\alpha} \frac{d \alpha}{d \eta} -1 \right) \frac{d m_\cK}{d \eta} \Biggl(4C_1m_\cK^3  - \frac{4 C_2}{m_\cK^5 R^8} \Biggr)\Biggr\} \approx 0, \label{EuL-eq1}
\\
 &\frac{4 c^2_\cK}{(2 \pi)^4 R^2} \frac{m_\cK}{\alpha^3} \Biggl[-2 \frac{d^2 \alpha}{ d\eta^2}- 5 \frac{d \alpha}{ d\eta} - 3 \alpha \Biggr]+\frac{\alpha^2 \cosh^3 \eta}{R^2}  \Biggl[2 d_\cK \Biggl(\frac{2}{\alpha} \frac{d \alpha}{ d\eta} \frac{d m_\cK}{ d\eta}+ 3 \frac{d m_\cK}{ d\eta} + \frac{d^2 m_\cK}{ d\eta^2}\Biggr)
\nn \\ 
  &+ 4 F_\cK^2 R^2 \frac{m^3}{\alpha^2}\Biggr]+\frac{\pi^2 }{2(2\pi)^4} \frac{1}{\alpha^4} \left[1-\frac{1}{\alpha} \frac{d \a}{d \eta}+ \left(\frac{1}{\alpha} \frac{d \a}{d \eta}\right)^2 \right]  \Biggl( 4C_1m_\cK^3  - \frac{4 C_2}{m_\cK^5 R^8} \Biggr)  \approx 0 \label{EuL-eq2}
\end{align}
in the late-time regime. Notice that we have dropped the contributions coming from $\mathcal{L}_m$ for brevity.

\subsection{Stabilization of $m_\cK$ and equation of motion for $\alpha$}

\begin{figure}[bht!]
\centering
\includegraphics[scale=0.82]{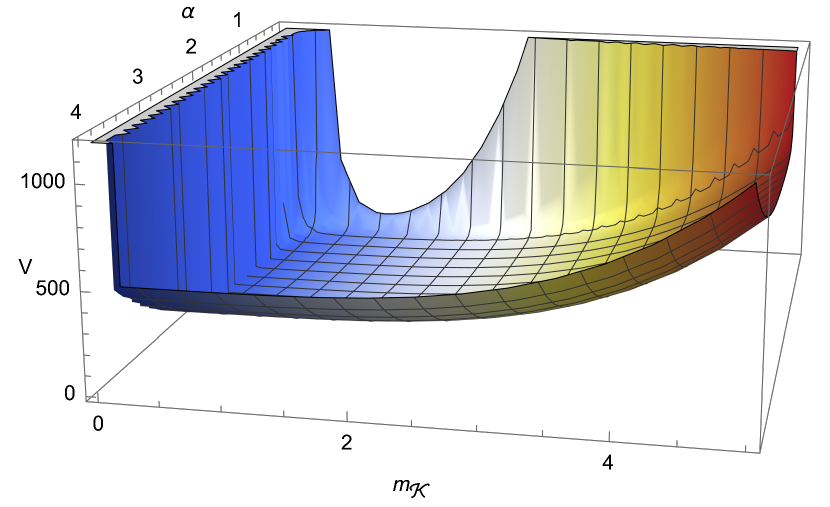}\vspace{0.3cm}
\caption{The effective potential \eqref{pot-cov-1} with all constants   set to unity in the early-time regime ($\eta=0.1$).}
\label{Fig-potential-1}
\end{figure}

\begin{figure}[bht!]
\centering
\includegraphics[scale=0.92]{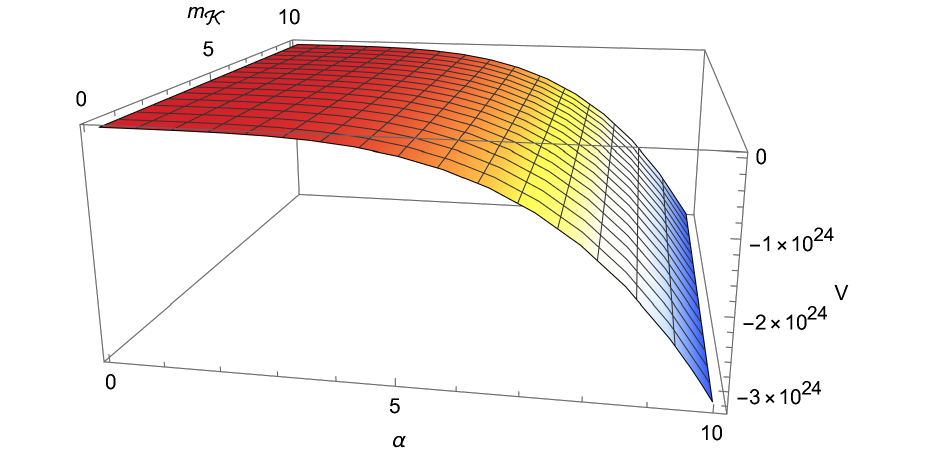}\vspace{0.3cm}
\caption{The effective potential \eqref{pot-cov-1} with all constants   set to unity in the late-time regime ($\eta=10$). }
\label{Fig-potential-2}
\end{figure}

Now consider the effective potential $V(\a, m_\cK)$ ,
which can be read off 
from the Lagrangian  \eqref{Lagrangian-grav-YM-vac}  written in the form $\cL = T - V$: 
\begin{align}
V(\a, m_\cK)&= 
\frac{V_3}{(2\pi)^4}\Biggl\{- \frac{6 c_{\mathcal{K}}^2}{ R^2} \frac{m_\cK^2}{\a^2} 
+ \frac{\pi^2}{2} \frac{1}{\alpha^4} \Biggl[\Big(C_1 + 32\pi^2 F^2_{\cK} \alpha^4 \cosh^3 \eta \Big) m_\cK^4 + \frac{C_2}{R^8 m_\cK^4 }  \Biggr] 
\nn \\
&+  \frac{\pi^2}{2} \Biggl[ \frac{C_3}{R^4} \frac{1}{\alpha^4} + \frac{64 \pi^2 N}{R^2 \tilde{R}^2} \alpha^4 \cosh^3 \eta \left(2-\cosh^2 \eta \right)\Biggr]+O(\varepsilon) \Biggr\} \ . 
\label{pot-cov-1}
\end{align}
The plots of $V(\a, m_\cK)$ both in the early-time and the late-time regimes are given in Figs. \ref{Fig-potential-1} and \ref{Fig-potential-2}, respectively. We observe that the potential attains a stable minimum for small $\eta$, but not for large $\eta$. 

\paragraph{Stabilization of $\cK$ and $\alpha$.}

At early times i.e. for sufficiently small $\eta$, we can 
assume that the vacuum energy is dominant with
$C_{1,2} \gg 0$, and dominates the contribution from the Einstein-Hilbert term $\sim\frac{m_\cK^2}{\a^2}$ as well as the YM term.
Dropping the $O(\varepsilon)$ terms,
this potential has a stable minimum 
for $m_\cK$ at  
\begin{align}
     m_\cK^8
    = \frac{1}{R^8} \frac{C_2}{C_1 +  32\pi^2 F^2_{\cK} \alpha^4  \cosh^3 \eta }.
\end{align}
This means that $\cK$ is indeed stabilized by quantum effects, which is an important result justifying the present scenario for emergent gravity, cf. Ref. \cite{Steinacker:2023myp}. 
%We also assume for simplicity that the contribution from $F_\cK^2$ is sub-leading and can be dropped; then $m_\cK^2 \approx const$, which will be assumed for simplicity. 
Moreover looking at Fig.  
\ref{Fig-potential-1}, it is manifest that the potential $V(\a, m_\cK)$
 has a stable local minimum for both variables, leading 
\begin{align}
 \alpha  \approx const \ 
\end{align}
at early times.
Therefore the undeformed $\cM^{3,1}$ background \eqref{undeformed-background} is consistent with quantum effects for early times.

Assuming $m_\cK = const$ more generally (which holds for sufficiently large $C_1/F_\cK^2 \gg 1$),
the Lagrangian functions \eqref{Lagr-grav-1}--\eqref{L-vac-1}  simplify as 
\begin{align}
\mathcal{L}_{\rm grav} & \approx  \frac{6 V_3 c_{\mathcal{K}}^2}{(2\pi)^4 R^2} \left(\frac{m_\cK }{\alpha }\right)^2  \left[1+\frac{5}{3 \alpha} \frac{d \alpha}{d \eta} + 2 \left(\frac{1}{\alpha}\frac{d \alpha}{d \eta}\right)^2  \right] + \Gamma_{\rm h.o.}(\cM,\cK),
\label{Lagr-grav-2}
 \\
 \mathcal{L}_{\rm YM} &\approx  \frac{V_3}{R^2} \alpha^2 \cosh^3 \eta \Biggl\{-F^2_{\cK} R^2  \frac{m^4_{\cK}}{\alpha^2}    - \frac{2N }{\tilde R^2 } \a^2   \left[2 - \left(1 + \frac{1}{\alpha} \frac{d \alpha}{ d \eta}\right)^2 \cosh^2\eta\right]\Biggr\},
 \label{Lagr-YM}
 \\
 \mathcal{L}_{\rm vac}   &\approx -\frac{\pi^2 V_3}{2(2\pi)^4} \frac{1}{\alpha^4}\left[1-\frac{1}{\alpha} \frac{d \a}{d \eta}+ \left(\frac{1}{\alpha} \frac{d \a}{d \eta}\right)^2 \right] \Biggl( C_1 m_\cK^4 + \frac{C_2}{m_\cK^4 R^8 } + \frac{C_3}{R^4} \Biggr), 
 \label{L-vac-2}
\end{align}
where recall that the first term
in $ \mathcal{L}_{\rm YM}$ arises from $\cK$, and the second term  from the $\cM^{3,1}$ background.
We can already recognize that at late times $\eta \gg 1$, the YM contribution from the $\cM^{3,1}$ background will be dominant. 
Then $\a$ will no longer be constant, which means that the late time evolution is modified by quantum effects. We will now examine this regime.

\paragraph{Late-time regime and accelerated expansion.}

Using Eqs. \eqref{Lagr-grav-2}--\eqref{L-vac-2}, the  Euler-Lagrange equation    for $\alpha$ becomes
\begin{align}
\mathscr{A}   \frac{d^2 \alpha}{ d\eta^2} + \mathscr{B} \left(\frac{d \alpha}{ d\eta}\right)^2 + \mathscr{C} \frac{d \alpha}{ d\eta} + \mathscr{E} \approx 0 \ ,
\label{alpha-equation-late-1}
\end{align}
where 
\begin{subequations}
\label{alpha-equation-parameter}
\begin{align}
 \mathscr{A} &:= 24 c^2_\cK \frac{m_\cK^2}{\alpha^4} + \frac{4 N (2 \pi)^4 }{\tilde{R}^2} \alpha^2 \cosh^5\eta -\frac{\pi^2 R^2}{\alpha^6}\Biggl(  C_1 m_\cK^4 + \frac{C_2}{m_\cK^4 R^8 } + \frac{C_3}{R^4} \Biggr),
\\
\mathscr{B} &:=-48 c^2_\cK \frac{m_\cK^2}{\alpha^5} + \frac{4 N (2 \pi)^4 }{\tilde{R}^2} \alpha \cosh^5\eta +\frac{3\pi^2 R^2}{\alpha^7}\Biggl(  C_1 m_\cK^4 + \frac{C_2}{m_\cK^4 R^8 } + \frac{C_3}{R^4} \Biggr),
\\
\mathscr{C} &:= \frac{20N (2 \pi)^4}{ \tilde{R}^2} \alpha^2 \cosh^5 \eta, 
\\
\mathscr{E} &:= 12 c^2_\cK \frac{m_\cK^2}{\alpha^3} + \frac{12N (2 \pi)^4}{ \tilde{R}^2} \alpha^3 \cosh^5 \eta \Biggl(1+ \frac{11}{6 \cosh^2 \eta}\Biggr)  -2\pi^2 R^2 \frac{1}{\alpha^5} \Biggl(  C_1 m_\cK^4 + \frac{C_2}{m_\cK^4 R^8 } + \frac{C_3}{R^4} \Biggr), 
\end{align}
\end{subequations}
which can be derived also from  Eq. \eqref{EuL-eq1} assuming  $m_\cK$  constant.

%\hcs{maybe can get first-order equations (corresponding to first Friedmann equations) from conserved e-m tensor!}

Let us consider the leading late-time terms occurring in Eq. \eqref{alpha-equation-late-1}. This means that we retain in Eq. \eqref{alpha-equation-parameter} only the terms proportional to $\cosh \eta \gg 1$, and we obtain the  equation 
\begin{align}
\alpha   \frac{d^2 \alpha}{ d\eta^2} +  \left(\frac{d \alpha}{ d\eta}\right)^2 + 5 \alpha \frac{d \alpha}{ d\eta} +3 \alpha^2 \approx 0 \ .
\end{align}
Upon using  the initial conditions
\begin{align}
\alpha(\eta=0)&= \alpha_0\ ,
\nonumber \\
\frac{d}{d \eta} \alpha(\eta=0) &=0
\end{align}
this leads to the solution 
\begin{align}
\alpha (\eta) \approx  \sqrt{3} \alpha_0 \, e^{-3 \eta/2} \, \sqrt{e^{ \eta} - 
\frac{2}{3}}\ ,
\label{alpha-full}
\end{align}
which can be further approximated at late times as 
\begin{align}
 \alpha (\eta) \approx   \sqrt{3} \alpha_0 \, e^{-\eta} \ .
 \label{alpha-eta-sol-2}
\end{align}
This is consistent with the classical solution\footnote{
This is not in conflict with the results in Ref. \cite{Sperling:2019xar}, where
a non-vanishing mass $m^2 > 0$ was added to the model.} derived in Appendix \ref{Appendix-C} for  $m^2=0$, see Eq. \eqref{class-background-m=0}. Notice that the exponential factor in Eq. \eqref{alpha-eta-sol-2}  leads to $\varepsilon \sim -1$ at late times.

We have thus found the one-loop corrected FLRW background of the model.
Identifying the effective metric for this background with time-dependent $\a$ is not trivial, because the $\hs$ components of the frame cannot be neglected, as discussed in section \ref{sec:linearized}. The effective  metric can be obtained by going to  adapted local normal coordinates, as discussed in  appendix \ref{sec:loc-normal-coords}. This leads to the effective metric \eqref{eff-metric-App}
\begin{align}
G_{\mu\nu} 
 = \sinh \eta \, (1 + \varepsilon\coth\eta)\a^{2} \eta_{\mu\nu} \  \sim  e^{-2\eta} \eta_{\mu\nu} \ ,
 \label{G-eff-accel-1}
\end{align}
%%V2 fixed 
using Eq. \eqref{alpha-full} at late times, which can be written as
\begin{align}
d s^2_G &= \sinh \eta \, (1 + \varepsilon)\a^{2} \eta_{\mu\nu} dx^\mu dx^\nu \
\sim \ -R^2 d \eta^2 
  + R^2 d \Sigma^2,
\label{naive-projected-G}
\end{align}
%%V2 
in hyperbolic coordinates using Eq. \eqref{eff-metric-FRW}.
By comparing with the standard FLRW metric \eqref{FRW-standard-metric}.
this is recognized as a $k=-1$ FLRW metric with $t \sim \eta$ and
constant scale parameter:
%\begin{subequations}
%\begin{align}
%dt^2 &\approx  R^2 (1+\varepsilon)\alpha^2 \sinh^3 \eta \, d \eta^2,  
%\label{dt-d-eta-rough-calc}
%\\
%a(t)^2 &\approx R^2 (1+\varepsilon) \alpha^2 \sinh\eta\cosh^2\eta 
%\label{a-squared-rough-calc}
%\end{align}    
%\end{subequations}
%at late times. Together with $\a \sim e^\eta$  this leads to $t \sim c e^{5 \eta/2}$, and  the cosmic scale parameter is obtained as
\begin{align}
   a(t) \sim const
  %% V2 is above   a(t) \sim \frac{5}{2} t
    \label{at-modified-NC-1}
\end{align}
at late times.
%corresponding to an asymptotically coasting FLRW cosmology.
The dilaton is found to be \eqref{dilaton-eta}
\begin{align}
     \r \sim  e^{-t} \ .
\end{align}
This should be compared with the undeformed metric \eqref{eff-metric-and-A} with $\a=const$, which leads to
an expanding behavior as $a(t) \sim \frac 32 t$  and increasing dilaton \cite{Sperling:2019xar}.

 We conclude that the present one-loop corrected background features an early expanding phase where $T^a \sim t^a$ is stabilized by the one-loop effective potential, and a late-time phase where 
$T^a \sim e^{-\eta} t^a$ is approaching the  solution of the classical Yang-Mills action without mass, cf. Eq. \eqref{class-background-m=0}. 
The underlying assumptions and their shortcomings will be discussed below.

A technical remark on the equations of motion is in order.
One might object that we have only considered the variations w.r.t. the cosmic scale parameter in deriving the equations of motion, and dropped all other possible variations of the background.
However this is sufficient for our purpose, because we are considering the most general $SO(3,1)$-invariant background\footnote{Recall that the most general $SO(3,1)$-invariant ansatz is given by Eq. \eqref{FLRW-frame--1}, where  we can assume $f=0$ by a suitable choice of gauge, as explained in Appendix \ref{sec:app-normal-gauge}.}, and $SO(3,1)$ is an exact symmetry of the quantum effective action. 
The general equations of motion at one loop for generic geometries will be derived elsewhere \cite{Kaushal}.

\section{Discussion and conclusions}\label{Sec:conclusions}

This paper should be seen as a step towards a better understanding of the classical and quantum dynamics of cosmological space-time branes in matrix models. 

The first important observation is that the 1-loop quantum corrections corresponding to vacuum energy and the 
induced Einstein-Hilbert term  significantly affect the cosmic evolution only in the early stage of the universe, 
where the vacuum energy leads via the potential $V(\a,m_\cK)$ to a stabilization of the classical $\cM^{3,1} \times \cK$ background. 
This justifies the use of $\cM^{3,1}$ as a consistent background, without having to introduce a mass term to the model by hand\footnote{Note that such a mass term breaks the crucial supersymmetry of the IKKT model, which is essential for the finiteness of the one-loop effective action.} as in earlier works \cite{Sperling:2019xar,Battista2022a}.

However at late times $\eta \gg 1$, the semiclassical YM action is found to dominate over the quantum effects, at least at one loop. 
%This is due to the overall $\cosh^3\eta$ factor in \eqref{Lagr-YM} as compared with the vacuum energy and the induced gravity term. 
Then the lack of stabilization by a mass term (or other possible mechanisms) leads to an exponential decay of the pre-factor $\a(\eta)$ of 
the generalized background \eqref{FLRW-background-alpha}, which approaches the classical solution of the model at late times with a constant scale factor $a(t) = const$, and decreasing dilaton. 

Clearly the late-time behavior of this cosmological solution is not satisfactory from a physical perspective. However, the present treatment is over-simplified, and should not be considered as definitive answer. The main simplification is in the assumption that the fuzzy extra dimensions $\cK$ are constant. This is certainly wrong, since $\cK$ is part of the dynamical background, whose evolution must be taken into account. 
One particularly intriguing scenario is that $\cK$ is stabilized by \qm{rotating} in the extra dimensions; note that this does not break any of the symmetries in a FLRW geometry. 
This will be studied in detail elsewhere.

There are several other aspects 
which need a more refined treatment. For example, a non-constant and decreasing dilaton will have important implications; 
in particular it implies that the Yang-Mills coupling  increases in time, which means that some quantum effects should in fact become stronger, notably those associated with $\cK$. This suggests that the present treatment is not complete and needs to be refined. We have also assumed that the constants $C_1, C_2$ and $C_3$ are all positive, which is not evident a priori; all these aspects require more detailed investigations.

Finally, to obtain a more complete understanding of the physics on these cosmological backgrounds, one must of course include matter and study the local perturbations. 
We have argued that matter is {\em not} expected to significantly affect the late-time evolution of the universe in the present framework, in  contrast to general relativity. However, a more complete understanding can only be obtained once a practical form of modified Einstein-like equations is available for the present framework; this will be addressed in future work.

%\subsection{The case of $k=0$ geometry}
 
%Now consider 
%\begin{align} 
 %  T^0 &= \a(x^0) t^0 + \b x^0   \nn\\
 %  T^i &= \g(x^0) t^i\nn\\
 %%   \tilde x^0 &= f(x^0) x^0 + g(x^0) t^0  \nn\\
  %  \tilde x^i &= h x^i 
%\end{align}
%Now there should be more than enough possibilities to obtain an invariant frame!

\section*{Acknowledgements}

Useful discussions with R. Brandenberger, S. Brahma, J. Nishimura, H. Kawai and other participants of the workshop \qm{Large-$N$ matrix models and emergent geometry} at the Erwin-Schr\"odinger institute for mathematics and physics in Vienna as well at EISA Corfu are gratefully acknowledged.
This work was supported by the Austrian Science Fund (FWF) grants P32086 and P36479.

\appendix
\numberwithin{equation}{section}

\section{$SO(3,1)$ invariant  gauge transformations and normal form}
\label{sec:app-normal-gauge}

For any given background $T^a$ of the form \eq{FLRW-frame--1}, the $x^a$ components of the 
background can be eliminated by a local $\L(\eta)$ gauge transformation that respects $SO(3,1)$.
Consider the infinitesimal gauge transformation
\begin{align}
    \d_\L = \{\L(x^4),.\} \ 
\end{align}
corresponding to a $\hs$-valued diffeo which respects the cosmic time.
It acts as
\begin{align}
    \d_\L (\a x^a) &=  R\tilde R^2 \a\L'\ t^a,  \nn\\
 \d_\L (\b t^a) &= R^{-1} \b\L'\  x^a,
\end{align}
for $a=0,...,3$
i.e. it rotates the $x^a$ and $t^a$ components.
This can be done independently for any time $\eta$, hence we can use such time-dependent $SO(3,1)$ gauge transformations to rotate the background into the form
\begin{align}
    T^a = \a(\eta) t^a \ 
    \label{FLRW-background-alpha-app}
\end{align}
denoted as {\em standard FLRW gauge}.

\section{Local normal coordinates and FLRW metric}
\label{sec:loc-normal-coords}

Consider again the background \eq{FLRW-background-alpha}, i.e., $T^a = \alpha(\eta) t^a$, leading to the $\hs$-valued frame \eq{frame-FLRW-hs-x}, which we write here again for convenience
\begin{align}
\label{frame-hs-app}
    E^{a\mu} =  \{T^a,x^\mu\} 
    &= \sinh \eta \,  \a \eta^{a\mu} 
+ \tilde R^2 R \a' t^a t^\mu.
\end{align}
 It is not possible to find 
 Cartesian normal coordinates $\tilde x^\mu$ which are manifestly covariant under the spacelike isometry $SO(3,1)$ such that the $\hs$ components of the frame vanish.  This appears to be a rather generic feature of geometries with curvature in the present framework.
 However, we can find such local (!) normal coordinates  if we give up the manifest $SO(3,1)$ invariance:

\paragraph{Local $SO(3)$-covariant normal coordinates.}

We are looking for $SO(3)$-invariant Cartesian isotropic local  coordinates $\tilde x^\mu$ near some reference point $\xi = (x^0,0,0,0)$, such that
the frame has no $\hs$ components at $\xi$, and satisfies
\begin{align}
   \tilde E^{a\mu}|_\xi :=  \{T^a,\tilde x^\mu\}|_\xi
 \stackrel{!}{=} \sinh \eta \,  \a(\eta) \eta^{a\mu} \ .
\end{align}
Since $x^i = 0$ ($i=1,2,3$) at $\xi$, only the spacelike components of the original frame have  $\hs$ components, and $E^{0i}|_\xi = 0 = E^{i0}|_\xi$. We  can thus assume 
\begin{align}
   \tilde x^0 &= x^0, \nn\\
  \tilde x^i  &= \phi^i_j(t,\eta) x^j, \qquad (i,j=1,2,3),
\end{align}
and the above requirement becomes
\begin{align}
     \{T^a,\tilde x^i\}|_\xi
    &= \phi^i_j E^{aj}|_\xi 
    \stackrel{!}{=} \sinh \eta \, \a\d^{ai} , \qquad a=1,2,3,   \nn\\
 \{T^0,\tilde x^i\}|_\xi
    &= \phi^i_j E^{0j}|_\xi = 0, \nn\\
\{T^i,x^0\}|_\xi
    &= E^{i0}|_\xi = 0,  \nn\\
\{T^0,x^0\}|_\xi
    &= E^{00}|_\xi = -\sinh \eta \,  \a.
\end{align}
Only the first equation is nontrivial, and it is solved by the ansatz
\begin{align}
     \phi^i_j  = \d^i_j + b t^i t_j \ . 
\end{align}
Then the first equation gives 
\begin{align}
    \left( \d^i_j + b t^i t_j \right)
    \left(\sinh \eta  \, \a \d^{aj} 
+ \tilde R^2 R \a' t^a t^j \right)  &=  \sinh \eta \,  \a\d^{ai}, 
%\sinh(\eta) \a \d^{ai}  + \tilde R^2 R \a' t^a t^i
% + b (\sinh(\eta) \a  t^i t_a + \tilde R^2 R \a' t^a  t^i (t_jt^j))   &= \sinh(\eta) \a\d^{ai} \nn\\
%\sinh(\eta) \a \d^{ai}  + \big(\tilde R^2 R \a'
% + b (\sinh(\eta) \a  
%+ R \a' \cosh^2\eta) \big) t^a  t^i  &= %\sinh(\eta) \a \d^{ai} 
\end{align}
%using $t\cdot t = \tilde R^2 \cosh^2(\eta)$,
which reduces using $t\cdot t = \tilde R^{-2} \cosh^2\eta$  (cf. Eq.   \eqref{t-mu-t-mu}) to
\begin{align} 
    \tilde R^2 R \frac{\a'}{\a} + b \left(\sinh \eta  \, 
    + R \frac{\a'}{\a} \cosh^2\eta\right) &= 0,
\end{align}
i.e.
\begin{align}
    b(\eta) = -\frac{\tilde R^2}{1
    + \varepsilon\coth\eta}\,
    \frac{\varepsilon}{\sinh\eta\cosh\eta} 
    \ \sim - \frac{\varepsilon}{1+\varepsilon}\frac{\tilde R^2}{\sinh^2\eta}
\end{align}
at late times.
This leads to the  local normal coordinates
\begin{align}
\tilde x^0 &= x^0,   \nn\\
    \tilde x^i &= x^i + b t_j x^j t^i 
    = x^i - b x^0 t_0 t^i \ .
    \label{LNC-FLRW-explicit}
\end{align}
These extend to a timelike observer near $x^i = 0$ for any time $x^0$, and
satisfy the desired property
\begin{align}
    \tilde E^{a\mu}|_\xi  = \sinh(\eta) \a\eta^{a\mu} \ 
\end{align}
as well as (cf. Eq. \eqref{x-mu-x-mu-prod})
\begin{align}
    \tilde x^i |_\xi &= 0 \ , \qquad \tilde x^\mu \tilde x_\mu |_\xi = x^\mu x_\mu = - R^2 \cosh^2\eta \ .
    \label{tilde-x-relations}
\end{align}
We can thus use the $\tilde x^\mu$ as local Cartesian coordinates to describe the local physics near $\xi$ for any $x^0$.
However,  the relation \eqref{A-rhoM-div} arising from the divergence constraint cannot be assumed in local normal coordinates, since
the $\hs$ components of the torsion are not negligible. We should therefore determine the dilaton directly using   Eq. \eqref{rho-M-G-relation}
\begin{align}
   \tilde \r^2  = \tilde\r_M^{-1}\sqrt{|G|} = \tilde\r_M \det \tilde E^{a\mu},
\end{align}
where $\tilde\r_M$ is the reduced symplectic density on $\cM^{3,1}$ in $\tilde x^\mu$ coordinates.
Note that the symplectic volume form $\Omega = \r_M d^4 x \Omega_t = \tilde\r_M d^4 \tilde x \Omega_t$ on $\C P^{1,2}$
(here $\Omega_t$ is the normalized volume form on the internal $S^2$) is rigid and not affected by the deformation of the frame. We can thus compute $\tilde\r_M$ using
\begin{align}
    \det\left(\frac{\del\tilde x^\mu}{\del x^\nu}\right) &= 1 + b t^\mu t_\mu
     = 1 + b \tilde R^{-2} \cosh^2\eta 
   %   = 1 -\frac{\varepsilon \coth \eta}{1 + \varepsilon\coth\eta}
    = \frac{1}{1 + \varepsilon\coth\eta} \ .
\end{align}
Recalling that $\rho_M = \frac 1{\sinh\eta}$ in Cartesian coordinates $x^\mu$,
the symplectic density on $\cM^{3,1}$ is therefore given by 
\begin{align}
    \rho_M d^4 x = \frac{1 + \varepsilon\coth\eta}{\sinh\eta} d^4 \tilde x =: \tilde \r_M d^4 \tilde x
\end{align}
in the $\tilde x^\mu$ coordinates,
with 
\begin{align}
    \tilde \r_M = \frac{1 + \varepsilon\coth\eta}{\sinh\eta} \ .
\end{align}
Therefore 
\begin{align}
    \tilde\r^2  = \tilde\r_M \det \tilde E^{a\mu}
     = (1 + \varepsilon\coth\eta)(\sinh\eta)^3 \a^4. 
\end{align}
Then the effective metric at $\xi$ is 
\begin{align}
\tilde G^{\mu\nu} = \tilde\r^{-2} \eta_{ab}\tilde E^{a\mu} \tilde E^{b\nu} 
 = \left(\sinh \eta \right)^{-1} (1 + \varepsilon\coth\eta)^{-1}\a^{-2} \eta^{\mu\nu}, 
\label{eff-metric-App}
\end{align}
in the above local normal coordinates.
For $\a\sim e^{-\eta}$ we have $\varepsilon \sim -1$, so that this metric  behaves like
\begin{align}
  \tilde G_{\mu\nu}\sim e^{-2\eta}\eta_{\mu\nu}  \ ,
\end{align}
at late times; recall that this metric and the underlying normal coordinates apply near $x^i=0$  for any (late) times. The dilaton is found to be 
\begin{align}
     \tilde\r^2 \sim  e^{-2\eta} \ .
     \label{dilaton-eta}
\end{align}
Since the relation 
$\tilde x^\mu \tilde x_\mu = - R^2 \cosh^2\eta$ given in Eq. 
\eqref{tilde-x-relations} still holds, 
this looks exactly like a $k=-1$  FLWR geometry \eqref{Frame-and-A} for such a 
local comoving observer, so that we can 
identify the scale parameter $a(t)$ as in Eq. \eqref{at-modified-NC-1}.

\section{Classical  solution for $m = 0$}
\label{Appendix-C}

In this section, we find a $k=-1$ FLRW solution
of the form \eqref{FLRW-frame--1}
\begin{align}
T^a &= \a(x_4) t^a \ 
\label{FLRW-frame--1a}
\end{align}
of the classical IKKT model without mass term,
with manifest $SO(3,1)$ symmetry.
Since the classical action dominates at late times, this is expected to be an approximate solution of the one-loop effective action at late times, dropping $\cK$ for simplicity;
%\begin{align}
%   - \Box_T T_b &= \{T^a,\{T_a,T_b\}\}
%= \{\a t^a + \b x^a,\{\a t_a + \b x_a,\a t_b + \b x_b\}\}
%\end{align}
recall that the simple $\cM^{3,1}$ background \eqref{undeformed-background} assumes an explicit mass term, which was
added to the model by hand for simplicity in Ref. \cite{Sperling:2019xar}.
To compute 
$\Box_T T_b$, we can use 
\eqref{TT-CR-general}
\begin{align}
   -\cF^{ab} &= 
    \{\a t^a,\a t^b\} 
 = -\frac{1}{\tilde R^2 R^2}\a^2\theta^{ab}  
 - \frac 1R \a \a'(x^a t^b - t^a x^b)
\end{align}
%V2 changed sign see (79)
and 
\begin{align}
     \{x^a t^b - t^a x^b,t_b\}
      &= - \frac{x^4}{R}t^a 
      + \frac{4 x^4}{R}t^a 
      +\frac{1}{R^2 \tilde R^2} \theta^{ab} x_b 
      =  \frac{4 x^4}{R}t^a, \nn\\
    \{x^a t^b - t^a x^b,x_b\}
      &= \frac{4x^4}{R} x^a + \theta^{ab} t_b
      - \frac{x^4}{R}x^a  = \frac{4x^4}{R} x^a, \nn\\
%    \{x^a t^b - t^a x^b,x_4^2\} &= -\frac{8x^4}{R} x^a x_a
%     = 8 R x^4 \big(1+ \frac{x_4^2}{R^2}\big) \nn\\
%    \{x^a t^b - t^a x^b,x_4\} &= 4 R \big(1+ \frac{x_4^2}{R^2}\big)
 \{x^a t^b - t^a x^b,x_4\} &= 0 \ ,
\end{align}
hence 
\begin{align}
    \{x^a t^b - t^a x^b,\a t_b\}
      &= \frac{4x^4}{R} \a t^a, \nn\\
       \{\theta^{ab},\a t_b\} 
      &= 3 \tilde R^2\a t^a \ .
\end{align}
This gives
\begin{align}
   - \Box_T T_b &= \{\a t^a,\{\a t_a,\a t_b\}\}  \nn\\
&= -\frac{1}{\tilde R^2 R^2} \{\a t^a,\a^2\theta^{ab}\} %% changed sig below
 - \frac 1R  \{\a t^a,\a \a'(x^a t^b - t^a x^b)\} \nn\\
%&= -\frac{1}{\tilde R^2 R^2} \big(\a^2 \{\a t^a,\theta^{ab}\} +\a \{ t^a,\a^2\}\theta^{ab}    \big)  \nn\\
% changed sign below
% & - \frac 1R \big(\a \a' \{\a t^a,(x^a t^b - t^a x^b)\}  + \a  \{t^a,\a \a'\} (x^a t^b - t^a x^b) \big) \nn\\
%&= -\frac{1}{\tilde R^2 R^2} 
%\big( 3\tilde R^2 \a^3 t^b -\frac 1R \a (\a^2)' x^a   \theta^{ab}    \big)  \nn\\
% V2 changed sign below
% & - \frac 1R \big(4\frac{x^4}{R}\a^2 \a'  t^b   -\frac 1R \a  (\a \a')' x^a   (x^a t^b - t^a x^b) \big) \nn\\
%&= -\frac{1}{R^2} \big( 3\a^3 + x_4\a (\a^2)' \big) t^b 
%V2 changed sign below
%  - \frac 1{R^2} \big(4x^4\a^2 \a'   + \a (\a \a')' (R^2 + x_4^2)\big)t^b \nn\\
&= - \frac {\a}{R^2} \left[6 x_4 \a \a'   + (\a \a')' (R^2 + x_4^2)  +  3\a^2 \right]t^b \nn\\
%  (old below)
%&=  \frac {\a}{R^2} \left[4x_4\a \a'  - 2 x_4 \a \a'   + (\a \a')' (R^2 + x_4^2)  -  3\a^2 \right]t^b \nn\\
   &\stackrel{!}{=} 0 \ ,
\end{align}
which vanishes  for 
\begin{align}
 6x_4\frac{\a'}{\a}   + \frac{(\a \a')'}{\a^2} (R^2 + x_4^2)
  +3 = 0 \ .
\end{align}
For late times $x_4 \gg R$, 
this  has two  asymptotic solutions: A) $\a(x_4) \sim x_4^{-1} = \frac 1{R\sinh \eta}$,
%where $\a'\a = -\frac{1}{x_4^3}$ and $\frac{\a'}{\a} = -x_4^{-1}$, hence 
%\begin{align}
% -6 + 3x_4^{-4} x_4^2 (R^2 + x_4^2)
%  +3 = 0 \ .
%\end{align}
corresponding to a background
\begin{align}
T^a \sim \frac 1{x_4} t^a \ 
\label{class-background-m=0}
\end{align}
consistent with Eq. \eqref{alpha-eta-sol-2}, or 
B) $\a(x_4) \sim x_4^{-3/2}$. 
However, both solutions are not really satisfactory: In case A), the frame becomes degenerate, and it is no longer possible to go to local coordinates. This problem disappears upon taking into account the quantum corrections, as discussed in section \ref{sec:eom-full}.
In case B), the 
frame is non-degenerate, but the
cosmic scale factor $a(t) \sim const$ would be constant  rather than expanding,
while the dilaton $\rho$ would approach zero.
This suggests that $\a(x^4)$ is not determined by the classical action, but significantly modified e.g. by quantum effects as considered in the main text.
Indeed, it is plausible that a consistent balance between classical and quantum effects require that $\rho \sim const$ at late times, which corresponds to $\a(x_4) \sim x_4^{-3/4}$. This will be studied in more detail elsewhere.

%This leads to the frame \eqref{frame-FLRW-hs-x} with
% $\varepsilon = O(1)$, hence with significant 
%non-negligible higher-spin components.
%As discussed in Appendix \ref{sec:loc-normal-coords}, these  
%$\hs$ components can be removed at any given (comoving) point
%in suitable local normal coordinates, where the 
%effective geometry is recognized as $k=-1$ FLRW cosmology 
%with $a(t) \sim \frac 52 t$.

%\begin{align}
%  0 = 2x_4^2  + x_4^2 -  3x_4^2
%\end{align}
%and the exact solutions are 
%\begin{align}
%    
%\end{align}

%We can then find (generalized) normal coordinates $\tilde x^\mu = x^\mu + \b t^\mu$
%such that 
%\begin{align}
%    - \Box_T \tilde x^\mu &= \g(x_4) \tilde x^\mu
%\end{align}
%and  therefore $\Box_T f(\tilde x) \in \tilde C^0$ are functions propagating in 3+1 dimensions.
%(they are not local normal coordinates in the strict sense since higher spin modes may not be preserved).

%This implies that we can extract an effective metric!

\bibliographystyle{JHEP}
\bibliography{references}

\end{document}